\documentclass[10pt,twocolumn,oneside,a4paper,final]{IEEEtran}
\usepackage{indentfirst}
\usepackage{setspace}
\usepackage[numbers,sort&compress]{natbib}
\usepackage{graphicx}
\usepackage{amstext}
\usepackage{algorithm}
\usepackage{algorithmic}
\usepackage{mathrsfs}
\usepackage{amssymb}
\usepackage{amsmath}
\usepackage{epstopdf}
\usepackage{CJK}
\usepackage{multicol}
\usepackage{stfloats}
\usepackage{url}
\usepackage{color}
\usepackage{enumerate}
\usepackage{diagbox}
\usepackage{tabularx}
\usepackage{footnote}
\makesavenoteenv{tabular}
\allowdisplaybreaks[0]
\ifCLASSINFOpdf
\else
\fi
\begin{document}
\title{Towards Large Intelligent Surface (LIS)-based Communications}

\author{Jide Yuan,~\IEEEmembership{Member, IEEE},
Hien Quoc Ngo,~\IEEEmembership{Senior Member, IEEE}, \\
and Michail Matthaiou,~\IEEEmembership{Senior Member,~IEEE}%

\thanks{A conference version of this paper appeared at IEEE ICC 2020 \cite{YuanConf}.}
%\thanks{Manuscript received September 17, 2019; revised February 20, 2020 and June 05, 2020; accepted July 07, 2020.}
%\thanks{W. H. Chin is with the Telecommunications Research Laboratory, Toshiba Research Europe, Bristol BS1 4ND, U. K. (e-mail: $\rm woonhau.chin@toshiba-trel.com$).}
%\thanks{This work was supported by the RAEng/The Leverhulme Trust Senior Research Fellowship LTSRF$1718\backslash14\backslash2$. The work of H. Q. Ngo was supported by the U.K. Research and Innovation Future Leaders Fellowships under Grant MR/S017666/1. The work of M. Matthaiou was supported by a research grant from the Department for the Economy Northern Ireland under the US-Ireland R\&D Partnership Programme. J. Yuan, H. Q. Ngo, and M. Matthaiou are with the Institute of Electronics, Communications and Information Technology (ECIT), Queen's University Belfast, Belfast, BT3 9DT, U.K. (e-mail: $\rm y.jide, hien.ngo, m.matthaiou@qub.ac.uk$).
%}
}

\maketitle

\begin{abstract}

The concept of large intelligent surface (\textbf{LIS})\textbf{-based} communication has recently raised research attention, in which a LIS is regarded as an antenna array whose entire surface area can be used for radio signal transmission and reception. To provide a fundamental understanding of \textbf{LIS-based} communication, this paper studies the uplink (UL) performance of LIS-based communication with matched filtering. We first investigate the new properties introduced by LIS. In particular, the array gain, spatial resolution, and the capability of interference suppression are theoretically presented and characterized.
Then, we study two possible LIS system layouts in terms of UL, i.e., centralized LIS (C-LIS) and distributed LIS (D-LIS).
Our analysis showcases that a centralized system has strong capability of interference suppression; in fact, interference can nearly be eliminated if the surface area is sufficient large or the frequency band is sufficient high.
For D-LIS, we propose a series of resource allocation algorithms, including user association scheme, orientation control, and power control, to extend the coverage area of a distributed system.
Simulation results show that the proposed algorithms significantly improve the system performance, and even more importantly, we observe that D-LIS outperforms C-LIS in microwave bands, while C-LIS is superior to D-LIS in mmWave bands. These observations serve as useful guidelines for practical LIS deployments.

\end{abstract}

\begin{IEEEkeywords}
Achievable spectral efficiency (SE), large intelligent surface (LIS), orientation control, power control, user association.
\end{IEEEkeywords}

% \fi
\IEEEpeerreviewmaketitle
\newtheorem{Definition}{Definition}
\newtheorem{Lemma}{Lemma}
\newtheorem{Theorem}{Theorem}
\newtheorem{Corollary}{Corollary}
\newtheorem{Proposition}{Proposition}
\newtheorem{Remark}{Remark}
\newtheorem{Property}{Property}
\newcommand{\rl}[1]{\color{red}#1}

%\section{Introduction}
%\IEEEPARstart{T}{he} introduction text goes here.

%\hfill January 11, 2007

\section{Introduction}

To support a diverse variety of applications including enhanced mobile broadband (eMBB), ultra reliability low latency communications (uRLLC) and massive machine-type communications (mMTC), an innovative concept that promotes the current state-of-the-art in wireless communications is urgently needed \cite{abs190210265,ATZORI20102787}.
Among various technologies recently proposed, an entirely new concept, namely, \emph{large intelligent surface (LIS)}, in which a spatially continuous surface is being used for signal transmission and reception, has attracted increasing attention \cite{BJORNSON2019}.
{In LIS systems, different from traditional massive multiple-input multiple-output (mMIMO), which integrates a vast amount of standard antenna elements in arrays, a large number of new-form antenna modules are deployed into a limited aperture, which forms a spatially continuous surface.}
As envisioned in \cite{8319526}, with a radically new design, a LIS has great ability to manipulate electromagnetic waves, and can theoretically make the entire wireless communication environment intelligent \cite{cui2017}.

Currently, the applications of the LIS concept are mainly divided into two categories: {(i) \textbf{LIS-assisted} wireless communications in which the LIS is regarded as a passive reflecting surface \cite{8811733,8644519,8683663,8466374}; and (ii)  \textbf{LIS-based} wireless communications in which the entire surface of the LIS is used for transmission and reception \cite{8108330}.}
%One is to consider a \textbf{LIS-assisted} wireless communication in which the LIS is regarded as a passive reflecting surface \cite{abs181003961,8644519,8683663,8466374}. Another is considered as \textbf{LIS-based} wireless communication architecture, in which the entire surface of LIS is used for transmission and reception instead of reflection \cite{8108330}.
For the former architecture, LIS comprises a large number of low-cost, programmable reflecting elements with an ``intelligent'' controller. By adjusting the phase shifts of each reflecting element independently according to the propagation conditions, the LIS has the capability of overcoming many fundamental limitations of radio propagation, e.g., blockage diffraction, and enhance the signal strength by aligning the phases of different paths or offering anomalous reflections.
In this context, the authors in \cite{nature2014} designed a tunable LIS-like architecture as a spatial microwave modulator.
The indoor simulation results showed that the signal strength between two antennas is enhanced by an order of magnitude. Therefore, thanks to the low power consumption of the passive reflection elements, the LIS can achieve higher data rate in a more energy-efficiency manner.
As a reflecting surface, LIS can be densely deployed around devices and terminals, which makes the propagation channel more line-of-sight (LoS) favorable, while a huge amount of overhead for channel state information (CSI) can be saved compared with conventional mMIMO \cite{abs190400453}. This is due to the fact that for a LoS-dominated channel, the channel components are highly correlated among LIS elements, making the overall channel matrix estimation possible by only a small number of channel sensors over the surface area.
In \cite{abs190410136}, a compressed sensing based channel estimation algorithm was proposed, through which the overall channel matrix can be reconstructed from the estimated channel obtained by active elements.
However, prior works on \textbf{LIS-assisted} wireless communication are mainly discussing the role of LIS for single-user scenarios. For multiuser scenarios, a passive reflecting surface can hardly adjust the phase shifts of each element to beamform to all users simultaneously \cite{8741198}.

{On the other hand, the \textbf{LIS-based} communication concept can be regarded as an extension of traditional mMIMO.
Apart from offering all the functionalities of traditional mMIMO, the \textbf{LIS-based} systems exhibit the following two new features.}
One is that, as the man-made structure of LIS enables to transmit and receive signals through its surface, the transmission power can be significantly reduced compared with traditional mMIMO.
As such, \textbf{LIS-based} systems can achieve the same level of performance with less power \cite{abs190400453}.
The other is that, in contrast to traditional antenna arrays, where the actual physical structures determine the radiation pattern of the signal, the LIS structure can control the electromagnetic field on the entire surface \cite{8319526}.
{Considering the physical limits of electromagnetic wave propagation, a practical implementation of LIS can be a compact integration of miniaturized antenna modules connected with software-controlled reconfigurable networks.
The purpose of this architecture is to harness the ability of the surface to manipulate the electromagnetic waves.
It has been demonstrated that the electromagnetic waves can be controlled by changing the coding sequences of ``0'' and ``1'' in real time, making the surface programmable \cite{liu870}.}
In sharp contrast to conventional arrays, where mutual coupling in radiation patterns occurs if the antenna spacing is less than half a wavelength,
thanks to recent advances in metamaterials, the distortions in the radiation patterns of LIS can be remarkably mitigated for any antenna spacing \cite{abs190308127}.
It is important to note that, different from \textbf{LIS-assisted} communications, \textbf{LIS-based} systems are capable of serving multiple users as transmission and reception can be performed across the entire surface.

For this reason, and in order to meet the massive connectivity requirements of the next generations of communication systems, the concept of \textbf{LIS-based} communications seems to have better scalability and fits better into the new wireless ecosystem. Therefore, this paper aims to look for a system layout that can fully deliver the potential of \textbf{LIS-based} communications.
Note that, to date, very few works have appeared that study the performance of \textbf{LIS-based} systems \cite{8319526,8264743,8108330,8580675}. This concept was firstly proposed in \cite{8108330}, in which the uplink (UL) rate was evaluated for an indoor scenario. The result shows a novel feature, namely that the multiplexing capability of \textbf{LIS-based} system is essentially determined by the wavelength $\lambda$.
Further, as extensions of \cite{8108330}, the authors studied the potential of LIS for positioning, in which the Cram\'{e}r-Rao lower bound for positioning a terminal on the central perpendicular line was derived in closed-form \cite{8264743}.
In \cite{8647606}, the capacity of \textbf{LIS-based} system was analyzed in the presence of hardware impairments, in which the impairments are modeled as a Gaussian process following the model of \cite{7080890}. An important observation is that the degradation of capacity caused by hardware impairments can be greatly suppressed by two approaches, i.e., enlarging the surface area and splitting a large LIS into a number of small LIS-units.
The authors in \cite{8948323} then investigated the feasibility of the latter approach in which each LIS-unit has a limited area. The asymptotic UL rate were given for massive users scenarios, which shows that a \textbf{LIS-based} system can achieve a comparable performance with conventional mMIMO with a significantly reduced area for antenna deployment.
However, the existing literature assumes that users are closely located around the LIS, i.e., the analysis based on this assumption is valid for near-field propagation environments, while the extension of this concept to far-field scenarios, the more common scenarios, is currently missing.

To give a full picture of \textbf{LIS-based} communication concept, this paper studies the system from an UL perspective with match-filtering (MF) for the far-field case. Specifically, the main contributions of this paper are summarized as follows.
\begin{itemize}
  \item
  We study the new properties introduced by LIS architecture. The results reveal the fact that array gain, and spatial resolution of a LIS architecture are highly dependent on the LIS size, orientation and frequency band.
  \item
  %With the new features introduced by LIS, the system performance of \textbf{LIS-based} communication will be entirely different from the conventional MIMO system. Therefore,
  We investigate the behavior of \textbf{LIS-based} communication in a multiuser scenario with two different layouts, i.e., centralized LIS (C-LIS) and distributed LIS (D-LIS), and aim to design effective schemes to maximize the sum spectral efficiency (SE) or maximize the minimum SE.
  \item
  For C-LIS, we first extend our LoS analysis to Ricean fading channels. Then, by utilizing the new properties introduced by LIS, we consider a brute-force searching for the sum SE maximization, in which the search space is effectively reduced down to the orientation domain, which largely reduces the algorithmic complexity.
  %For C-LIS, by utilizing the observing the fact that the interference is significantly eliminated with a larger surface area or at higher frequency bands, we proposed a brute-force searching the search space can be reduced down to the orientation domain, which largely reduces the complexity.
  %as the per-user rates are coupled among users forming a non-convex rate maximization problem, a brute-force searching is the only method that can solve the problem. Fortunately,
  \item
  For D-LIS, we study the system performance under the assumption that the each of the distributed LIS-unit serves one particular user. To fully reap the potential of D-LIS, we propose a series of algorithms to rationally allocate and schedule resources, including a large scale fading (LSF)-based user association scheme, an orientation control (OC) algorithm, and a max-min power control (PC) algorithm.
  \item
  We numerically demonstrate that the proposed resource allocation algorithms can significantly boost the system performance in terms of both the sum SE and minimum user SE.
  %More importantly, with the help of the proposed resource allocation algorithms, D-LIS offers a much higher coverage probability compared with C-LIS.
\end{itemize}

The rest of this paper is organized as follows:
We present the system model in Section II, and evaluate the new properties of LIS architecture in Section III. The performance analysis and the corresponding resource allocation algorithms of C-LIS and D-LIS are analyzed in Section IV and Section V, respectively. Numerical results are presented in Section VI, and our main observations are summarized in Section VII. Proofs are relegated to Appendices.

\textbf{Notation}---Throughout this paper, vectors and matrices are denoted in bold lowercase letters and bold uppercase letters, respectively.
The complex and real number fields are represented by $\mathbb{C}$ and $\mathbb{R}$, respectively.
%We use the notation $x\sim\mathcal{CN}(0,1)$ to denote that $x$ is Gaussian distributed with zero mean and unit variance.
We use $\left\{\mathbf{A}\right\}_{i}$ and $a_{i,j}$ to denote the $i$th row and $\left(i,j\right)$th entry in matrix $\bf{A}$, respectively.
The operation $\left\|{\bf A}\right\|_p$ denotes the $p$-norm of the matrix $\bf A$.
The superscripts $\left(\cdot\right)^*$, notation ${\mathsf{E}}\left[\cdot\right]$ and notation $\mathsf{var}$ denote the Hermitian conjugate, the expectation, and variance, respectively. %Finally, $(\bf A \cdot B)$ denotes the Hadamard product (element-wise product between matrices $\bf A$ and $\bf B$), and ${\bf A}^{\circ (a)}$ denotes an element-wise exponentiation operation on $\bf A$, i.e., $\left({\bf A}^{\circ (a)}\right)_{i,j}=\left({\bf A}_{i,j}\right)^a$.

\section{System Model}

Consider a two-dimensional circular\footnote{We consider a circular LIS for the sake of mathematical tractability. Note, however, that for the considered case of far-field propagation, the shape of the LIS is not important as the LIS can still be regarded as a continuous surface of fixed area \cite{8264743}.} LIS deployed on the $xy$-plane with radius $R$, and $K$ single-antenna users located in a three-dimensional space. For ease of understanding, we first consider a typical scenario to investigate the fundamental properties, in which the LIS center is located at $x=y=z=0$, while the users are located at the space $z>0$. We assume a far-field propagation scenario where the path loss between a particular user to every point on the LIS is the same. Specifically, we consider the distance between the $k$th user located at $\left(x_k,y_k,z_k\right)$ to the LIS center as the effective distance, which is given by
\begin{equation}\label{d_c_k}
{d_{k}^c} = \sqrt {{x_k^2} + {y_k^2} + {z_k^2}}.
\end{equation}
The far-field free-space path loss is then expressed as a function of the distance between the transmit and receive antennas \cite{molisch2004ieee}
%The free space attenuation has different property with respect to the distance, and can be classified into near-field and far-field channel. Specifically, the value of free-space path loss (FSPL) in near-field and far-field can be evaluated as functions of distance between antennas, and are given by
%\begin{equation}\label{}
%{\rm{P}}{{\rm{L}}_{{\rm N},k}} ={\left( {\frac{{\rm{1}}}{{{\rm{2}}\kappa {d_k^c}}}} \right)^{\rm{2}}}\left( {{\rm{1 + }}{{\left( {\frac{{\rm{1}}}{{\kappa {d_k^c}}}} \right)}^{\rm{2}}}} \right),
%\end{equation}
%and
\begin{equation}\label{PL_k}
{\text{P}}{{\text{L}}_{k}} = {\left( {\frac{{\rm{1}}}{{2\kappa {d_k^c}}}} \right)^2},
\end{equation}
where $\kappa  = \frac{2\pi }{{\lambda }}$ with $\lambda$ being the wavelength. Note that the far-field path loss is valid when $d_k^c $ is larger than the \emph{Fraunhofer distance}, i.e., $d_k^c > \frac{8R^2}{\lambda}$.

%Note that the near and far fields are defined in terms of the \emph{Fraunhofer distance}, and is given by the following
%\begin{equation}\label{}
%{d_{\rm{F}}} = \frac{{2{L^2}}}{\lambda },
%\end{equation}
%where $L$ is the largest dimension of LIS.

%\subsection{Channel Model}

%In the near-field radiating structure, as channel propagation is usually dominated by the LoS channel, we consider a perfect LoS propagation environment without any scatters.

%In near-field scenarios, the phase shift of channel for a user to the point at LIS is mainly dominated by the distance between them.
%Hence, the general channel propagation from the $k$th user to the point $(x,y,0)$ at LIS can be addressed as \cite{}
%%\begin{equation}\label{}
%%{h^{\rm n}_{{x_k},{y_k},{z_k}}}\left( {x,y} \right) = {\rm{P}}{{\rm{L}}_{k}^{\frac{1}{2}}} \cdot {e^{ - j\kappa {d_k}}},
%%\end{equation}
%\begin{equation}\label{}
%{h^{\rm n}_k}\left( {x,y} \right) = {\rm{P}}{{\rm{L}}_{k}^{\frac{1}{2}}} \cdot {e^{ - j\kappa {d_k}}},
%\end{equation}
%where
%%${\phi _k}\left( {x,y} \right) = {{{z_k}} \mathord{\left/
%%{\vphantom {{{z_k}} {{d_k}}}} \right.
%%\kern-\nulldelimiterspace} {{d_k}}}$ is angle-of-arrival (AoA) of the transmitted signal from the $k$th user to the point $(x,y,0)$ at LIS,
%${d_k}$ is the distance between the $k$th user and the point $\left(x,y,0\right)$ at LIS, and is given by
%\begin{equation}\label{}
%{d_{k}} = \sqrt {{z_k^2} + {{\left( {x - {x_k}} \right)}^2} + {{\left( {y - {y_k}} \right)}^2}}.
%\end{equation}
%The radiating model is depicted in Fig. 1(a).

\subsection{Channel Model}

%In far-field scenarios, due to the present of the scatters and reflects, we consider both the LoS-only and LoS-included channel, i.e., Ricean channel, in this paper.
%In this paper, we consider a LoS dominated channel. This assumption is sensible for two reasons.

We consider a LoS-dominated propagation environment.\footnote{{We first consider a LoS-dominated propagation environment, whilst the more general case of Ricean fading is considered later in the paper.}} {This LoS propagation model is reasonable for systems deployed in indoor or outdoor open spaces such as rural areas \cite{6362603,7973186}, and for millimeter wave wireless systems with very small cell sizes \cite{6932503}. In addition, the LISs are naturally deployed much higher above the sea level, e.g., on the top of buildings, making the signal strength from LoS path significantly larger than that from scattering paths. Even if the LoS component is blocked, there exist many scenarios, where a strong specular component dominates over the weak scattered components that can be neglected \cite{5374071}.}
%More importantly, as clarified in \emph{Remark} 2 later in this paper, with the MF applied at the LIS architecture, the effective channel from a LoS path and scattering path are nearly orthogonal due to the capability of interference suppression, which results in a LoS-dominated propagation environment.

For a far-field scenario, since the distance between a user and LIS is sufficient large, we assume that the angles-of-arrival (AoA) for a user to each point at the LIS are identical. Therefore, the general channel propagation from the $k$th user to the point $(x,y,0)$ at a typical LIS can be represented as
\begin{equation}\label{g_los_k}
{g_{k}}\left( {x,y} \right) = {\text{PL}}_k^{\frac{1}{2}} \cdot {h_{k}}\left( {x,y} \right),
\end{equation}
where
\begin{equation}\label{h_los_k}
{h_{k}}\left( {x,y} \right) = {e^{ - j\left( {\kappa {d_k} + {\varphi _k}} \right)}},
\end{equation}
%${\phi _k}\left( {x,y} \right) = {{{z_k}} \mathord{\left/
%{\vphantom {{{z_k}} {{d_k}}}} \right.
%\kern-\nulldelimiterspace} {{d_k}}}$ is angle-of-arrival (AoA) of the transmitted signal from the $k$th user to the point $(x,y,0)$ at LIS,
and $\varphi_k$ is the original phase of the $k$th user which follows uniform distribution in the range of $[-\pi,\pi]$,
${d_k}$ is the distance between the $k$th user and the point $\left(x,y,0\right)$ at the LIS, given by
\begin{equation}\label{d_k}
{d_{k}} = \sqrt {{z_k^2} + {{\left( {{x_k}-x} \right)}^2} + {{\left( {{y_k}-y} \right)}^2}}.
\end{equation}
It is important to note that minor differences between the distance from a particular user to any two points at the LIS will hardly impact the path loss but highly impact the phase of $h_k(x,y)$. For this reason, we treat the effect of distance on the path loss and phase shift separately.

%{\rl\subsection{Ricean Channel}
%
%Consistent with conventional channel model, the Ricean channel is given as
%%\begin{equation}\label{}
%%h_{{x_k},{y_k},{z_k}}^{{\text{R}}}\left( {x,y} \right) = \frac{1}{{\sqrt {\Omega  + 1} }}{\text{PL}}_k^{\frac{1}{2}} \cdot \left( {\sqrt \Omega  {e^{ - j\kappa \left( {d_k^c +\vartriangle d\sin {\phi _k}} \right)}} + a{e^{ - j\theta }}} \right),
%%\end{equation}
%%\begin{equation}\label{}
%%h_{{\rm R},k}\left( {x,y} \right) = \frac{1}{{\sqrt {\Omega  + 1} }}{\text{PL}}_k^{\frac{1}{2}} \cdot \left( {\sqrt \Omega  {e^{ - j\kappa \left( {d_k^c +\vartriangle d\sin {\phi _k}}+\varphi_k  \right)}} + a{e^{ - j\theta }}}\right),
%%\end{equation}
%\begin{equation}\label{g_r_k}
%{g_{{\text{R}},k}}\left( {x,y} \right)  = {\text{PL}}_k^{\frac{1}{2}}\cdot {h_{{\text{R}},k}},
%\end{equation}
%where
%\begin{equation}\label{h_r_k}
%{h_{{\text{R}},k}}\left( {x,y} \right)  = \frac{1}{{\sqrt {\Omega  + 1} }}\left( {\sqrt \Omega  {h_{{\text{Los}},k}}\left( {x,y} \right) + a_k{e^{ - j\theta_k }}} \right),
%\end{equation}
%where $\Omega$ is the Ricean factor. The second term represents the channel gain from all the possible scattering paths, in which $a_k$ is the overall attenuation which follows Rayleigh distribution
%\begin{equation}\label{Rayleigh}
%{f_{a_k}}\left( r \right) = 2r{e^{ - {r^2}}},
%\end{equation}
%and $\theta_k$ is the uniformly distributed phase shift which is uncorrelated to the signal.
%
%\textbf{This part may leave for future work.}}

\subsection{Effective Channel and Achievable SE with MF Scheme}

Based on (\ref{h_los_k}), the received signal at the LIS location $(x, y, 0)$ from all $K$ users is given by
\begin{equation}\label{}
r\left( {x,y} \right) = \sum\nolimits_{k = 1}^K {\sqrt {{p_k}} {g_{k}}\left( {x,y} \right){s_k}}  + n\left( {x,y} \right),
\end{equation}
where {$p_k$ is transmitted power}, $s_k$ is the transmitted signal of the $k$th user with $\left\|s_k\right\| = 1$, and $n\left( {x,y} \right)$ is the AWGN at LIS with variance ${\sigma ^2}$.
{We assume that MF is applied at the LIS thanks to its low complexity and the fact that it can be implemented in a distributed manner.}
With MF at the LIS, the received signal for the $k$th user is given as \cite{8319526}
\begin{align}\label{}
{r_k}\left( {x,y} \right)= \sum\nolimits_{k' = 1}^K {\sqrt {{\text{P}}{{\text{L}}_k}{\text{P}}{{\text{L}}_{k'}} {p_{k'}}}} {\Sigma^{\mathcal{S}} _{kk'}}{s_{k'}} + {\omega _{k}},
\end{align}
where the effective channel
\begin{align}\label{Coeffi_kk}
{\Sigma^{\mathcal{S}} _{kk'}} \triangleq \iint_{\left( {x,y} \right) \in {\mathcal{S}}} {h_{k}^ * \left( {x,y} \right){h_{k'}}\left( {x,y} \right)dxdy},
\end{align}
where ${\mathcal{S}}$ is the surface-area of the LIS; ${\omega _k}$ is the noise after MF with zero-mean and variance
\begin{equation}\label{noise_p}
{\mathsf{E}}\left[ {\omega _k^ * {\omega _{k}}} \right] = {\text{P}}{{\text{L}}_k}{\Sigma^{\mathcal{S}}_{kk}}{\sigma ^2}.
\end{equation}

With the effective channel given in (\ref{Coeffi_kk}) and the noise model in (\ref{noise_p}), the achievable SE of the $k$th user with surface area $\mathcal S$ is then calculated as
\begin{equation}\label{Rate_general}
{\mathsf{R}_k}  = {\log _2}\left( {1 + \frac{{{p_k}{\text{P}}{{\text{L}}_k}{{\left( {\Sigma _{kk}^\mathcal{S}} \right)}^2}}}{{\Sigma _{kk}^\mathcal{S}{\sigma ^2} + \sum\nolimits_{k' \ne k} {{p_{k'}}{\text{P}}{{\text{L}}_{k'}}{{\left| {\Sigma _{kk'}^\mathcal{S}} \right|}^2}} }}} \right).
\end{equation}
Note that for $K=1$, the achievable SE has a similar structure as the one in \cite{8647606}, which was obtained for single-user scenarios in near-field propagation environments.

\section{Intrinsic Properties of LIS Architectures}

%Ideally, LIS can realize transmission and reception through each point on the surface. Hence, the overall channel propagation from a particular user to LIS are determined by the shape and size of LIS. In this paper, we consider a continuous surface in circle shape.

%In addition, a more engineering-implement layout of LIS is to approximate the continuous surface via a dense deployed conventional antenna-array.
%The main drawback of this approach may be the serious autocorrelation across the antenna-array due to the insufficient distance between antennas.

\begin{figure}[!b]
	\centering
	\includegraphics[width=6cm]{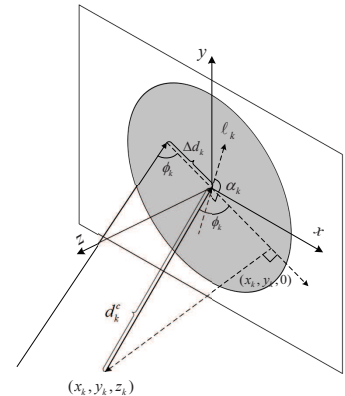}
	\caption{The radiating model of a transmitting signal to a circular LIS in far-field scenarios. {A zero-crossing $\ell_k$ is perpendicular to the projection line of the $k$th user's signal direction on the $xy$-plane. The channel phase shift can be evaluated by using the distance $\Delta d_k$ to the reference line $\ell_k$.}}
\end{figure}

In this section, we investigate the effective channel to illustrate the new features introduced by LIS architecture.

To evaluate the coefficient $\Sigma^{\mathcal{S}}_{kk'}$ with circular LIS, it is necessary to measure the wave phase at each point of the LIS. We first project the radiation direction of the $k$th user on $xy$-plane, as shown in Fig. 1.
%Since the AoAs $\phi_k$ for the $k$th user to each point at LIS are identically the same, and it is clear that for any lines that are perpendicular to the projection line of signal direction on $xy$-plane, the wave phases of each point on the line are also identical.
For any line that is perpendicular to the projection line of signal direction on the $xy$-plane, the wave phases of the points on this line are identical.
This comes from the fact that the AoAs for a particular user to the points at the LIS are the same.
Hence, denoting by $\alpha_k$ the angle of elevation (AoE), and consider the wave phase on the zero crossings line
\begin{equation}\label{line_k}
{\ell _k} = \left\{ {\left( {x,y} \right):x - y\tan {\alpha _k} = 0} \right\}
\end{equation}
as reference, the phase at point $(x,y,0)$ can be calculated via its distance to ${\ell _k}$. The channel response is then rewritten as
\begin{equation}\label{}
{h_{k}}\left( {x,y} \right) ={e^{ - j\left( {\kappa d_k^c + \Delta d_k\kappa \cos {\phi _k} + {\varphi _k}} \right)}},
\end{equation}
where
\begin{equation}\label{delta_k}
\Delta d_k = \frac{{y - x\tan {\alpha _k}}}{{\sqrt {{{\tan }^2}{\alpha _k} + 1} }}.
\end{equation}
For ease of understanding, we now define the following coefficients.

\begin{Definition}
We define ${\eta _{kk'}}$, ${\xi _{kk'}}$, and ${\zeta _{kk'}}$, which have the following form\footnote{As shown in (\ref{angle2cor1}) and (\ref{angle2cor2}) and with the fact that $\sin\phi_k=\frac{z_k}{d^{c}_k}$, all coefficients can be expressed as a function of AoA and AoE. The use of coordinates is for simplicity and intuitive.}
\begin{align}
\label{xi_kk}&{\eta _{kk'}} = {{{x_k}} \mathord{\left/
 {\vphantom {{{x_k}} {d_k^c}}} \right.
 \kern-\nulldelimiterspace} {d_k^c}} - {{{x_{k'}}} \mathord{\left/
 {\vphantom {{{x_{k'}}} {d_{k'}^c}}} \right.
 \kern-\nulldelimiterspace} {d_{k'}^c}},\\
\label{eta_kk}&{\xi _{kk'}} = {{{y_k}} \mathord{\left/
 {\vphantom {{{y_k}} {d_k^c}}} \right.
 \kern-\nulldelimiterspace} {d_k^c}} - {{{y_{k'}}} \mathord{\left/
 {\vphantom {{{y_{k'}}} {d_{k'}^c}}} \right.
 \kern-\nulldelimiterspace} {d_{k'}^c}},\\
\label{zeta_kk}&{\zeta _{kk'}} = {{{z_k}} \mathord{\left/
 {\vphantom {{{z_k}} {d_k^c}}} \right.
 \kern-\nulldelimiterspace} {d_k^c}} - {{{z_{k'}}} \mathord{\left/
 {\vphantom {{{z_{k'}}} {d_{k'}^c}}} \right.
 \kern-\nulldelimiterspace} {d_{k'}^c}},
\end{align}
%\begin{equation}\label{xi_kk}
%{\eta _{kk'}} = {{{x_k}} \mathord{\left/
% {\vphantom {{{x_k}} {d_k^c}}} \right.
% \kern-\nulldelimiterspace} {d_k^c}} - {{{x_{k'}}} \mathord{\left/
% {\vphantom {{{x_{k'}}} {d_{k'}^c}}} \right.
% \kern-\nulldelimiterspace} {d_{k'}^c}},
%\end{equation}
%\begin{equation}\label{eta_kk}
%{\xi _{kk'}} = {{{y_k}} \mathord{\left/
% {\vphantom {{{y_k}} {d_k^c}}} \right.
% \kern-\nulldelimiterspace} {d_k^c}} - {{{y_{k'}}} \mathord{\left/
% {\vphantom {{{y_{k'}}} {d_{k'}^c}}} \right.
% \kern-\nulldelimiterspace} {d_{k'}^c}},
%\end{equation}
%\begin{equation}\label{zeta_kk}
%{\zeta _{kk'}} = {{{z_k}} \mathord{\left/
% {\vphantom {{{z_k}} {d_k^c}}} \right.
% \kern-\nulldelimiterspace} {d_k^c}} - {{{z_{k'}}} \mathord{\left/
% {\vphantom {{{z_{k'}}} {d_{k'}^c}}} \right.
% \kern-\nulldelimiterspace} {d_{k'}^c}},
%\end{equation}
and
\begin{equation}\label{chi_kk}
{\chi^2 _{kk'}} = \eta _{kk'}^2 + \xi _{kk'}^2.
\end{equation}
\end{Definition}

%For a continuous surface, when use whole surface for transmission, the overall channel gain from a particular user to the LIS should be integrated over the surface $\mathcal S$. Thus, the overall LoS-Only channel can be calculated as
%\begin{equation}\label{}
%g_k^c = \int_{\Delta d \in \mathcal{S}} {h_{{x_k},{y_k},{z_k}}^{{\text{fLos}}}\left( {\Delta d} \right)d\Delta d}.
%\end{equation}
%the overall near-field channel can be calculated as
%\begin{equation}\label{}
%{g_k^c} = \iint_{x,y \in \mathcal{S}} {h^{\rm n}_{{x_k},{y_k},{z_k}}}\left( {x,y} \right) dxdy.
%\end{equation}
%Can't make it.

All the parameters ${\eta _{kk'}}$, ${\xi _{kk'}}$, ${\zeta _{kk'}}$ and ${\chi _{kk'}}$ are related to the difference of users' position in the spatial domain.
With the definitions above, we are ready to analytically evaluate the effective channel.

\begin{Proposition}
The effective channel $\Sigma^{\mathcal{S}}_{kk'}$ of a circular LIS equals{\footnote{In a far-field propagation environment, the variation of the AoAs from a user to each point at surface can be ignored. This is the reason that the effective channel in \emph{Proposition 1} is different from the result in \cite{8319526} for a near-field propagation environment.}}
\begin{equation}\label{}
{\Sigma^{\mathcal{S}}_{kk'}} = {{\rm A}_{kk'}} \cdot {\rm B} \left(R,\kappa,{\chi _{kk'}}\right),
\end{equation}
where
\begin{equation}\label{}
{{\rm A}_{kk'}} = {e^{j\left( {\kappa \left( {d_k^c - d_{k'}^c} \right) + {\varphi _k} - {\varphi _{k'}}} \right)}},
\end{equation}
\begin{equation}\label{}
{\rm B} \left(R,\kappa,{\chi _{kk'}}\right)= 2\pi R\frac{{{J_1}\left( {R\kappa {\chi _{kk'}}} \right)}}{{\kappa {\chi _{kk'}}}},
\end{equation}
where $R$ the is radius of the circle, whilst $J_1(\cdot)$ is the Bessel function of the first kind.
\end{Proposition}
\begin{IEEEproof}
See Appendix A.
\end{IEEEproof}

Note that ${{\rm A}_{kk'}}$ is a constant phase shift which depends on the users' positions and original phase, while ${\rm B} \left(R,\kappa,{\chi _{kk'}}\right)$ is the \emph{LIS response} with MF which reveals the interference suppression and the spatial resolution of the LIS with respect to its size.
We now further investigate the coefficient ${\Sigma}_{kk'}^{\mathcal{S}}$ to obtain more analytical insights.

\subsection{Array Gain}

Array gain relates to the received signal power at the LIS corresponding to the desired signal part \cite{8319526}. It can be defined as the effective channel gain when $k'=k$. We have the following property for the array gain.

\begin{Property}
$\Sigma^{\mathcal{S}}_{kk'}$ represents the array gain when $k'=k$, which equals to
%\begin{equation}\label{}
%{\rm {B}}\left(R,\kappa, {{\chi _{kk}}} \right) = \pi R^2
%\end{equation}
\begin{equation}\label{}
\Sigma^{\mathcal{S}}_{kk} = \pi R^2.
\end{equation}
\end{Property}
\begin{IEEEproof}
It is intuitive that ${\rm{A}}_{kk}=1$, and by leveraging \emph{L'Hospital's Rule}, we have
\begin{equation}\label{}
\mathop {\lim }\limits_{x \to 0} \frac{{{J_1}\left( {ux} \right)}}{x} = \mathop {\lim }\limits_{x \to 0} \frac{{\partial {J_1}\left( {ux} \right)}}{{\partial x}} \mathop  = \limits^{(a)} {\left. {\frac{u}{2}\left( {{J_0}\left( {ux} \right) - {J_2}\left( {ux} \right)} \right)} \right|_{x = 0}},
\end{equation}
where $(a)$ is obtained from \cite[Eq. 03.01.20.0006.01]{123456789}. Noting that $J_0(0)=1$ and $J_2(0)=0$, we complete the proof.
\end{IEEEproof}

%\begin{figure}[!b]
%	\centering
%	\includegraphics[width=9.5cm]{B_C_R.pdf}
%	\caption{The coefficient $\left|{\Sigma}_{kk'}^{\mathcal{S}}\right|$ with respect to $\chi_{kk'}$ for a circular LIS.}
%\end{figure}

The result holds for any circular LIS with finite surface-area.
The conclusion that the array gain equals to the surface area makes intuitive sense, which is, to some extent, similar to the conventional antenna array whose array gain in a LoS environment approaches to the number of elements in the antenna array \cite{1424539}.
%Clearly, since the channel components at different LIS points are highly correlated in LoS environment, the conventional assumption of antenna spacing of greater than $\lambda/2$ is not necessary.
%reveals an impressive gain compared to the conventional MIMO system.
%The work in \cite{1424539} shows that array gain of the conventional antenna array approaches to the number of antenna elements in antenna array. However, the result
%makes intuitive sense, which is, to some extent, similar to the conventional antenna array whose array gain approaches to the number of antenna elements in antenna array \cite{1424539}.
Fig. 2a shows the absolute value of the effective channel ${\Sigma}_{kk'}^{\mathcal{S}}$ with respect to $\chi_{kk'}$, in which the $\chi_{kk'}=0$ case represents the array gain of LIS. It can be observed that the array gain equals to the surface area as expected. {Note that the result in Fig. 2a has similar form as the result for the near-field scenario \cite{8319526}. Yet, they are mathematically quite different, since the results in \cite{8319526} are approximated by a $\mathsf{sinc}$ function, while our results are expressed through a Bessel function of the first kind.}
%A more remarkable insight is that with larger size of LIS, ${\Sigma}_{kk'}^{\mathcal{S}}$ converges more quickly with respect to $\chi_{kk'}$.
Moreover, with increasing $\chi_{kk'}$, $|{\Sigma}_{kk'}^{\mathcal{S}}|$ decreases in an oscillatory manner, and converges to zero, which indicates that the effective channels from users further apart are almost orthogonal.
However, as the surface area of LIS is limited in practice, it is of interest to investigate the spatial resolution of a LIS.

%This result demonstrates the interference suppression of the LIS with respect to its size. Similar conclusion has been drawn in \cite{8319526}, in which with sufficient large surface area of LIS, any two users can be almost separated without interference, even if they are located very close together.

\begin{figure*}[!t]
	\centering
	\includegraphics[width=16.5cm]{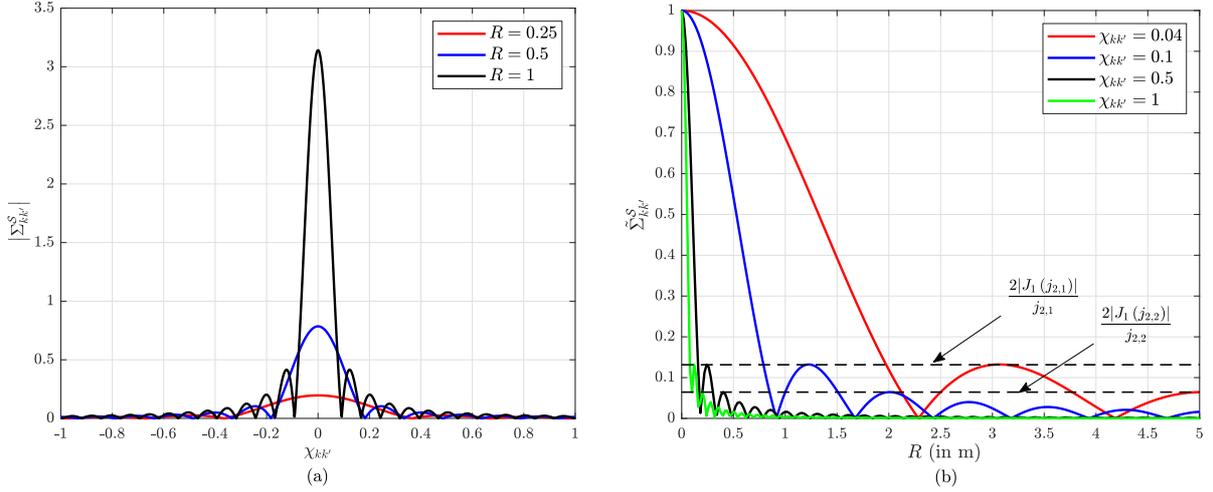}
	\caption{The effective channel for a circular LIS, in which (a) shows $\left|{\Sigma}_{kk'}^{\mathcal{S}}\right|$ with respect to $\chi_{kk'}$, and (b) compares ${\tilde{\Sigma}}_{kk'}^{\mathcal{S}}$ with respect to $R$.}
\end{figure*}

\subsection{Spatial Resolution}

The spatial resolution represents the minimum related distance of two users so that the ratio of the interference to the array gain is smaller than a predefined and small threshold. The spatial resolution is a very important characteristic. It helps us to determine how well two users can be separated with respect to their distance. A precise definition of the spatial resolution is as follows.
%We consider that, when $\frac{\left|{\rm B} \left({R,\chi _{{\rm Los},kk'}}\right)\right|}{{\rm B} \left({R,\chi _{{\rm Los},kk}}\right)}<\eta$, i.e., $\frac{\left|{\rm B} \left({R,\chi _{{\rm Los},kk'}}\right)\right|}{\pi R^2}<\eta$, LIS can separate the $k$th and the $k'$th users, where $\eta $ is the threshold with range $0<\eta \ll 1$. With which, we define the spatial resolution as follows.
\begin{Definition}
By denoting $\tilde\Sigma^{\mathcal{S}}_{kk'}=\left|\frac{\Sigma^{\mathcal{S}}_{kk'}}{\Sigma^{\mathcal{S}}_{kk}}\right|$ the normalized effective channel, %i.e., $ \tilde\Sigma^{\mathcal{S}}_{kk'}=\frac{\left|\Sigma^{\mathcal{S}}_{kk'}\right|}{\pi R^2}$,
the spatial resolution $\bar\chi$ is then defined as:
\emph{%Given a circular LIS with radius $R$,
for any two users $k$ and $k'$ whose $\chi_{kk'}>\bar\chi$, $\tilde\Sigma^{\mathcal{S}}_{kk'} <\eta$, where $\eta$ is a small positive value.}
\end{Definition}

We then state the following lemma that can be used to evaluate the spatial resolution.

%draw the channel respondence normalized by $\pi R^2$ with respect to the radius, shown in Fig. 3. We find that the first minima of all curves has same value

%To proof the \emph{Proposition} 2, we first investigate the property of following function.

\begin{Lemma}
The function $f(ux)=\frac{{{J_1}\left( {ux} \right)}}{{ux}}$ converges to zero with increasing $x$.
Specifically, the function exhibits damped oscillation, and each of the local minima and maxima is a constant value which is uncorrelated to scale parameter $u$, and is given by
$\frac{{{J_1}\left( {{j_{2,n}}} \right)}}{{{j_{2,n}}}}$, for $n \in {\mathbb{N}_ + }$, where $j_{m,n}$ is the $n$th zero of the $J_m(\cdot)$ function. Moreover, the absolute value of $\frac{{{J_1}\left( {{j_{2,n}}} \right)}}{{{j_{2,n}}}}$ decreases with respect to $n$.
\end{Lemma}
\begin{IEEEproof}
The convergence of the function is obvious by recalling the well-known feature that the radius of convergence of the Bessel function of the first kind is infinite. Then,
by differentiating the function with respect to $x$ in the range $x>0$, and let the result equal to zero, we have
\begin{equation}\label{}
\frac{\partial }{{\partial x}}\left( {\frac{{{J_1}\left( {ux} \right)}}{{ux}}} \right) \mathop  = \limits^{(b)}  - \frac{{{J_2}\left( {ux} \right)}}{x}=0,
\end{equation}
where $(b)$ is due to \cite[Eq. 03.01.20.0009.01]{123456789}. The roots of the equation are simply obtained as
${x_n} = \frac{{{j_{2,n}}}}{u}$, for $n \in {\mathbb{N}_ + }$.
Substituting ${x_n}$ into $f(ux)$, we obtain the formula of each maxima and minima. Along with the fact that the absolute value of local minima and maxima of $J_1(\cdot)$ monotonically decreases with $n$, we complete the proof.
\end{IEEEproof}

%\begin{figure}[!h]
%  \centering
%  \includegraphics[width=12.5cm]{B_C_chi_kk_woR.pdf}
%  \caption{The normalized coefficient ${\rm B}_{\rm {Los},kk'}$ with respect to $\chi_{{\rm Los},kk'}$ for a circle LIS.}
%\end{figure}

With the property given above, we can observe that $\left|f(ux)\right|$ cannot reach $\frac{{\left| {{J_1}\left( {{j_{2,n}}} \right)} \right|}}{{{j_{2,n}}}}$ again when $x> {{j_{2,n}}}$. This phenomenon perfectly matches the definition of $\bar\chi$ since when we choose $\frac{{\left| {{J_1}\left( {{j_{2,n}}} \right)} \right|}}{{{j_{2,n}}}}$ as $\eta$ and ${j_{2,n}}$ as $\bar\chi$, for any $x>j_{2,n}$, we have $|\tilde\Sigma^{\mathcal{S}}_{kk'}|<\eta$. Therefore, we obtain the spatial resolution criterion in the following proposition.

\begin{Proposition}
By setting $\eta_n = \frac{2{\left| {{J_1}\left( {{j_{2,n}}} \right)} \right|}}{{{j_{2,n}}}}$ as our threshold, the spatial resolution of a circular LIS with respect to the its radius is given as
${\bar\chi} = \frac{{{j_{2,n}}}}{{\kappa R}}$, for $n \in {\mathbb{N}_ + }$, where $n$ is adjustable according to the resolution requirements.
\end{Proposition}

The result can be directly obtained using \emph{Lemma} 1. From the above expression, we clearly observe that the spatial resolution increases with the carrier frequency, and the LIS size. For example, when setting the threshold $\eta_2 =\frac{2\left|{{J_1}\left( {{j_{2,2}}} \right)}\right|}{{{j_{2,2}}}} \approx 0.0645$, the spatial resolution $\bar\chi$ approximately equals to $\frac{1.3396}{R}\lambda$. Fig. 2b shows $\tilde\Sigma^{\mathcal{S}}_{kk'}$ with respect to $R$, which verifies our analysis. It can be seen that the absolute value of the normalized response monotonically decreases in an oscillatory manner. Moreover, with larger size, the LIS is able to obtain higher spatial resolution.
%From engineering perspective, spatial resolution provides the limits of the capability of the interference suppression for LIS, i.e., the interference caused by the user further apart than $\bar\chi$ can be nearly ignored.
Similar conclusion has been drawn in \cite{8319526}, in which with sufficient large surface area of the LIS, any two users can be almost separated without interference, even if they are located very close together.

%More specifically, we list several results for $\lambda = 0.15$\,m (i.e., 2\,GHz) in Table I for practical reference.
%\begin{table}[!h]
%\begin{center}
%\caption{Spatial Resolution $\chi_{kk'}$ with different $\eta_n$ and $R$.}
%\begin{tabular}{p{3cm}<{\centering}|p{1cm}<{\centering}|p{1cm}<{\centering}|p{1cm}<{\centering}|p{1cm}<{\centering}}
%  \hline
%  \hline
%  \diagbox{$\eta_n$}{$\bar\chi$}{$R$ (m)} & 0.5 & 1 & 2 & 4 \\
%  \hline
%  \hline
%  0.132 & 0.245 & 0.123 & 0.061 & 0.031\\
%  \hline
%  0.064 & 0.402 & 0.201 & 0.100 & 0.050\\
%  \hline
%  0.040 & 0.555 & 0.277 & 0.139 & 0.069\\
%  \hline
%  \hline
%\end{tabular}
%\end{center}
%\end{table}

\begin{Remark}
Different from conventional mMIMO, \emph{Proposition} 2 shows that the spatial resolution of the LIS can reach extremely high precision when the frequency is high, which implies that LISs have a strong capability of interference suppression at high frequency bands.
This new feature indicates that a LIS can theoretically create a nearly interference-free propagation environment, and thus, an impressive gain can be obtained at high frequency bands in the presence of a LoS channel.
\end{Remark}

\begin{figure}[!b]
	\centering
	\includegraphics[width=6.5cm]{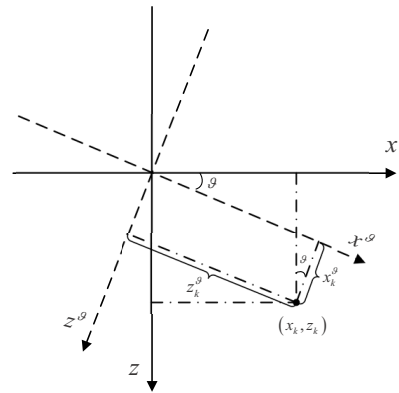}
	\caption{The schematic diagram of LIS unit on $xz$-plane.}
\end{figure}

%\begin{Remark}
%Assume that a scatterer is placed besides a user, and further assume that the signal from the LoS path and the reflected signal off the scatterer reach the LIS simultaneously. In this case, the reflected signal can be regarded as a LoS signal transmitted from an image of the user created by the scatterer with weaker strength. As illustrated in Fig. 2b, a LIS is capable of suppressing the interference from the other paths.
%Spatial resolution in \emph{Proposition} 2 provides the limits of the capability of the interference suppression for LIS, i.e., the interference caused by users further apart than $\bar\chi$ can be nearly ignored.
%%In general, the further apart the user and the scatterer are, and the larger surface area is, the weaker the interference. Therefore, with the MF procedure, the scattering signals can be viewed as orthogonal to the signal from the LoS path.
%\end{Remark}

%
%\begin{figure}[!t]
%	\centering
%	\includegraphics[width=9.5cm]{B_C_R_ratio.pdf}
%	\caption{The normalized coefficient ${\tilde{\Sigma}}_{kk'}^{\mathcal{S}}$ with respect to $R$ for a circular LIS.}
%\end{figure}

\subsection{Orientation Adjustable LIS}

With the result given above, the interference, or say, effective channel $\Sigma^{\mathcal{S}}_{kk'}$, is highly correlated with the coefficient $\chi_{kk'}$, which is determined by the AoAs of users (see \emph{Footnote 2} in page 8).
One of the advantages of LIS architectures is that they can intelligently control their orientation based on the AoAs of the received signal to minimize inter-user interference.
We, hence, evaluate the LIS response for an orientation adjustable LIS in this section.

We assume that the LIS can ideally adjust its angle $\vartheta$ in the range $[{-\pi},{\pi}]$ along the $y$-axis, as shown in Fig. 3; the normalized LIS response (with respect to $\vartheta$) is then given in the following proposition.

\begin{Proposition}
Denote by ${\rm \tilde B} \left(R,\kappa,{\chi ^{\vartheta}_{kk'}}\right)= \frac{{\rm B} \left(R,\kappa,{\chi ^{\vartheta}_{kk'}}\right)}{{\rm B} \left(R,\kappa,{\chi ^{\vartheta}_{kk}}\right)}$ the normalized LIS response for an orientation adjustable LIS. Then, ${\rm \tilde B} \left(R,\kappa,{\chi ^{\vartheta}_{kk'}}\right)$ with respect to $\vartheta$ is given as
\vspace{-0.1cm}
\begin{equation}\label{}
{\rm{\tilde B}}\left( {R,\kappa ,\chi _{kk'}^\vartheta } \right) = \frac{{2{J_1}\left( {R\kappa \chi _{kk'}^\vartheta } \right)}}{{R\kappa \chi _{kk'}^\vartheta }},
\end{equation}
where
\vspace{-0.1cm}
\begin{equation}\label{chi_vartheta}
\left(\chi _{kk'}^\vartheta\right)^2  = \xi _{kk'}^{\text{2}} + {{\left( {\eta _{kk'}^{{\vartheta }}} \right)}^2},
\vspace{-0.1cm}
\end{equation}
with
\vspace{-0.1cm}
\begin{equation}\label{eta2}
{\left( {\eta _{kk'}^\vartheta } \right)^2} = \eta _{kk'}^2{\cos ^2}\vartheta  + \zeta _{kk'}^2{\sin ^2}\vartheta  + 2{\eta _{kk'}}{\zeta _{kk'}}\cos \vartheta \sin \vartheta.
\vspace{-0.1cm}
\end{equation}
\end{Proposition}
\begin{IEEEproof}
See Appendix B.
\end{IEEEproof}

The coefficient ${\rm{\tilde B}}\left( {R,\kappa ,\chi _{kk'}^\vartheta } \right)$ reveals the capability of interference suppression of LIS as a function of its orientation.
Note that the result is identical to the one in \emph{Proposition} 1 when $\vartheta =0$.
As our goal is to manipulate the LIS to minimize interference, it is necessary to find the minimum value of ${\rm{\tilde B}}\left( {R,\kappa ,\chi _{kk'}^\vartheta } \right)$ in the range $[-\pi,\pi]$.

\begin{Property}
The range of the absolute value of ${\chi _{kk'}^{{\vartheta}}}$ with respect to ${\vartheta}$ is
\vspace{-0.1cm}
\begin{equation}\label{range_chi}
{\left| {\chi _{kk'}^{{\vartheta}}} \right|} \in \left[ {\left|\xi _{kk'}\right|, \left|\varpi_{kk'}\right| } \right],
\end{equation}
where $\varpi^2_{kk'} = \xi _{kk'}^{\text{2}} + \eta _{kk'}^2 + \zeta _{kk'}^2$.
Therefore, given a fixed LIS size and $\lambda$, the minimum value of the normalized channel response ${\rm \tilde B} \left(R,\kappa,{\chi ^{\vartheta}_{kk'}}\right)$ is
\begin{equation}\label{min_norm_B}
\min \left\{ {{\rm{0,\tilde B}}\left( {{R},\kappa ,\left|{\xi _{kk'}}\right|} \right),{\rm{\tilde B}}\left( {{R},\kappa ,\left|\varpi_{kk'}\right| } \right)} \right\},
\end{equation}
where ${\rm \tilde B} \left(R,\kappa,{\chi ^{\vartheta}_{kk'}}\right)=0$ iff ${\chi ^{\vartheta}_{kk'}}=\frac{j_{1,n}}{R\kappa}, n \in {\mathbb{N}_ + }$
exists in the range of $[\left|{\xi _{kk'}}\right|,\left|\varpi_{kk'}\right|]$,
and the corresponding $\vartheta$ is $\vartheta = \frac{1}{2}\arctan {\bar v}$,
where
\begin{align}\label{solve_v}
&\bar v= \frac{{\text{1}}}{{4\eta _{kk'}^2\zeta _{kk'}^2 - {c^2}}}\notag\\
&\times\left(\! { - {\eta _{kk'}}{\zeta _{kk'}}\left( {\eta _{kk'}^2\! - \!\zeta _{kk'}^2} \right) \pm c\sqrt {{{\left( {\eta _{kk'}^2 + \zeta _{kk'}^2}\! \right)}^2} \!-\! {c^2}} } \right),
\end{align}
with
\begin{equation}\label{}
c = 2{\left( {\frac{{{j_{1,n}}}}{{R\kappa }}} \right)^2} - \xi _{kk'}^{\text{2}} - \varpi _{kk'}^2.
\end{equation}
\end{Property}
\begin{IEEEproof}
See Appendix C.
\end{IEEEproof}

As $j_{1,n}$ is fixed by nature, given a $R$ and $\kappa$, we can simply check if there exists a zero point satisfying ${\chi ^{\vartheta}_{kk'}}=\frac{j_{1,n}}{R\kappa}$ in the range of $[\left|{\xi _{kk'}}\right|,\left|\varpi_{kk'}\right|]$. Otherwise, the phase shift $\vartheta$ that minimizes interference will be one of the solutions of
\begin{equation}\label{var_theta_solu}
\left\{ \begin{gathered}
  {{\hat \vartheta }_{1}} = \frac{1}{2}\arctan \frac{{2{\eta _{kk'}}{\zeta _{kk'}}}}{{\eta _{kk'}^2 - \zeta _{kk'}^2}}, \hfill \\
  {{\hat \vartheta }_{n}} = {{\hat \vartheta }_{1}} + \left(n-1\right)\frac{\pi }{2},\;\;\;n=2,3,4, \hfill \\
\end{gathered}  \right.
\end{equation}
as proved in Appendix C.

\begin{Remark}
{Compared with conventional antenna arrays, LISs provide an additional domain, i.e., orientation domain, to enhance and optimize the signal quality according to channel conditions.
By performing orientation control ahead of the MF process, interference can be significantly reduced.
This new characteristic offers a remarkable flexibility on scheduling, and enables LISs to fully reap their potential of interference suppression.}
\end{Remark}

%Without searching across range of $\pi$, it is straightforward to obtain that the overall computational complexity of the suboptimal solutions is ${\mathcal O}(K^2)$. Note that there is a very high chance to find a zero point in the given range as long as users are not locate extremely close.

%{\rl \begin{Proposition}
%The effective Ricean channel $\Sigma^{\mathcal{S}}_{{\text{R}},kk'}$ in channel with circle LIS equals
%\begin{align}\label{}
%{\Sigma^{\mathcal{S}}_{{\text{R}},kk'}} &= \frac{\Omega }{{\Omega  + 1}}{\Sigma^{\mathcal{S}} _{{\text{Los}},kk'}} + \frac{1}{{\Omega  + 1}}{a_k}{a_{k'}}{e^{j\left( {{\theta _k} - {\theta _{k'}}} \right)}} \hfill \notag\\
%&+ \frac{{\sqrt \Omega  }}{{\Omega  + 1}}\left( {{a_{k'}}{{\rm B}}\left(R, {{{\cos }}{\phi _k}} \right){e^{j\left( {\kappa d_k^c + {\varphi _k} - {\theta _{k'}}} \right)}} + {a_k}{{\rm B}}\left( R, {{{\cos }}{\phi _{k'}}} \right){e^{ - j\left( {\kappa d_{k'}^c + {\varphi _{k'}} + {\theta _k}} \right)}}} \right).
%\end{align}
%\end{Proposition}
%\begin{IEEEproof}
%See Appendix B.
%\end{IEEEproof}
%
%\textbf{This part may leave for future work.}}

\section{Performance Analysis of C-LIS}

%In this section, we study the achievable rate for a LIS system in different topology layouts, i.e., centralized LIS (C-LIS) system and distributed LIS (D-LIS) system. Our aim is to design effective schemes to maximize the sum rate and maximize the minimum rate for both architectures.
%To further exhibit the potential of D-LIS, we design a power control and an user association scheme based on the received signal strength.

%\subsection{Centralized LIS}
In this section, we study the achievable SE for a LIS system in centralized layouts, i.e., C-LIS system. The main benefit of a centralized system comes from the enhanced array gain and the powerful capability of interference suppression due to a larger surface area.

We consider a C-LIS system, in which the LIS can intelligently adjust its orientation and change the surface area that is being used for transmission.
The system performance of such an architecture is then assessed in the following proposition.

\begin{Proposition}
In a C-LIS, the achievable SE of the $k$th user for an orientation adjustable LIS is given as
\begin{equation}\label{R_k}
{\mathsf{R}_k} = {\log _2}\left( {1 + \frac{{{p_k}{\text{P}}{{\text{L}}_k}}}{{\frac{1}{{\pi {R^2}}}{\sigma ^2} + \sum\nolimits_{k' \ne k} {{p_{k'}}{\text{P}}{{\text{L}}_{k'}}{\rm{\tilde B}}{{\left( {R,\kappa ,{\chi^{\vartheta} _{kk'}}} \right)}^2}} }}} \right),
\end{equation}
and the overall sum SE across $K$ users then equals
\begin{equation}\label{R_k_sum}
{\mathsf{R}^{\text{C-LIS}}_{{\text{total}}}} = \sum\nolimits_{k = 1}^K {{\mathsf{R}_{{k}}}}.
\end{equation}
\end{Proposition}
\begin{IEEEproof}
The result can be directly obtained based on \emph{Proposition} 1 and (\ref{Rate_general}).
\end{IEEEproof}

The result in (\ref{R_k}) reveals the joint impact of the LIS radius, wavelength, angle $\vartheta$ and users' positions on the achievable SE, which indicates that LIS is able to maximize ${\mathsf{R}_k}$ by adjusting the frequency band, its size and its orientation.
To provide more insightful results, we now investigate two extreme cases.

\subsection{Large LIS or high frequency band}

When the size of LIS is sufficient large or the frequency band is sufficient high, according to \emph{Property} 2, it is intuitive that the channel response ${\rm{B}}\left( {R,\kappa ,{\chi^\vartheta _{kk'}}} \right)$ normalized by the array gain $\pi R^2$ satisfies
\begin{equation}\label{Interf_inf}
\mathop {\lim }\limits_{R \to \infty } {\rm{\tilde B}}\left( {R,\kappa ,{\chi^\vartheta _{kk'}}} \right) = \mathop {\lim }\limits_{\kappa  \to \infty } {\rm{\tilde B}}\left( {R,\kappa ,{\chi^\vartheta _{kk'}}} \right) = 0.
\end{equation}
The equation indicates that, for the $k$th user, the interference caused by other users can be almost canceled at high frequency bands or with a large LIS. Hence, we obtain the user sum SE in the following corollary.

\begin{Corollary}
In a C-LIS, if $R$ is sufficient large or/and the frequency is high, the achievable SE of the $k$th user can be approximated as
\begin{equation}\label{R_k_inf}
{\mathsf{R}_k}  \approx {\log _2}\left( {1 + \frac{{p_k}}{{{\sigma ^2}}}{\pi {R^2}}{\text{P}}{{\text{L}}_k}} \right),
\end{equation}
and the overall sum SE across $K$ users then equals
\begin{equation}\label{R_k_sum_inf}
{\mathsf{R}^{\text{C-LIS}}_{{\text{total}}}} \approx \sum\nolimits_{k = 1}^K {{{\log }_2}\left( {1 + \frac{{p_k}}{{{\sigma ^2}}}{\pi {R^2}}{\text{P}}{{\text{L}}_k}} \right)} .
\end{equation}
\end{Corollary}
%\begin{IEEEproof}
%The expression can be obtained directly by substituting \emph{Property} 1 and (\ref{Interf_inf}) into (\ref{Rate_general}).
%\end{IEEEproof}

%\begin{figure}[!h]
%	\centering
%	\includegraphics[width=12.5cm]{Rate_highf.pdf}
%	\caption{The comparison of achievable sum rate with respect to $\lambda$. The results are shown for $R=1$\,m, and are averaged over 100 runs.}
%\end{figure}

The result shows a similar conclusion as in conventional mMIMO, in which when the number of antennas at BS is infinite, the interference can be fully canceled.
Moreover, we observe an advantage of LIS compared with conventional mMIMO, which is the great capability of interference suppression at high frequency bands. In other words, the result in (\ref{R_k_sum_inf}) can be regarded as an upper bound of the achievable SE.

\begin{figure*}[!b]
	\centering \includegraphics[width=16.5cm]{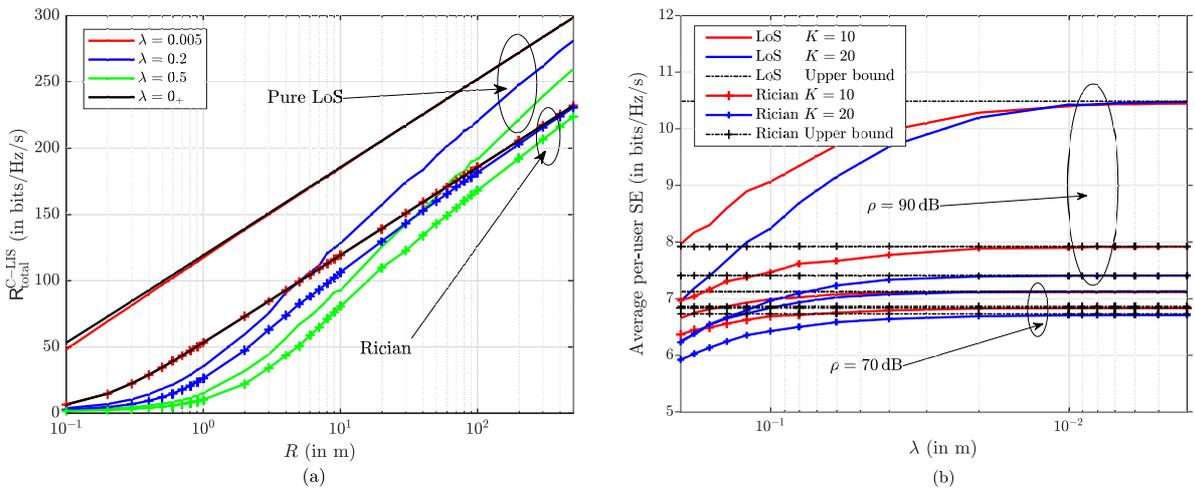}
	\caption{Performance of C-LIS, in which (a) compares the $\mathsf{R}^{\text{C-LIS}}_{\text{total}}$ against $R$ with $\rho = 100$\,dB, and (b) compares the per-user SE with respect to $\lambda$. The results are shown for $\gamma_{1,\ldots,K} = 10$\,dB and $\sigma^2 = -174$\,dBm/Hz/s, and averaged over 100 runs.}
\end{figure*}

{\subsection{Ricean fading channels}

In centralized systems, the propagation environment is more likely to experience Ricean fading instead of a pure LoS channel. In this case, the channel response from the $k$th user to the point $(x,y)$ at LIS can be modeled as
\begin{equation}\label{}
{{\tilde h}_k}\left( {x,y} \right) = \sqrt {\tfrac{\gamma_k }{{1 + \gamma_k }}} {h_k}\left( {x,y} \right) + \sqrt {\tfrac{1}{{1 + \gamma_k }}} {g_k}\left( {x,y} \right),
\end{equation}
where $\gamma_k$ represents the Ricean factor for the $k$th user, ${g_k}\left( {x,y} \right)\sim \mathcal{CN}(0,1)$ represents the NLoS channel from the $k$th to the point $(x,y)$.
The effective channel $\tilde \Sigma _{kk'}^\mathcal{S}$ can then be evaluated as in (\ref{tilde_Sigma}) at the top of next page,
\begin{figure*}[!t]
\vspace{-0.1cm}
\begin{align}\label{tilde_Sigma}
\tilde \Sigma _{kk'}^\mathcal{S} &= \iint_{\left( {x,y} \right) \in \mathcal{S}} {\tilde h_k^*\left( {x,y} \right){{\tilde h}_{k'}}\left( {x,y} \right)dxdy}\notag\\
& = \tfrac{1}{{\sqrt {\left( {1 + {\gamma _k}} \right)\left( {1 + {\gamma _{k'}}} \right)} }}\left( {\sqrt {{\gamma _k}{\gamma _{k'}}} \Sigma _{kk'}^\mathcal{S} + \sqrt {{\gamma _k}} {{\text{C}}_k} \cdot \Omega _{kk'}^\mathcal{S} + \sqrt {{\gamma _{k'}}} {{\text{C}}_k} \cdot \Omega _{kk'}^\mathcal{S} + D_{kk'}^\mathcal{S}} \right)
\end{align}
\vspace{-0.1cm}
\hrulefill
\end{figure*}
where ${{\text{C}}_k} = {e^{j\left( {\kappa d_k^c + {\varphi _k}} \right)}}$, ${{\text{C}}_{k'}}={e^{ - j\left( {\kappa d_{k'}^c + {\varphi _{k'}}} \right)}}$,
$\Omega _{kk'}^\mathcal{S}= \int_{\left( {x,y} \right) \in \mathcal{S}} {{e^{j\Delta {d_k}\kappa \cos {\phi _k}}}{g_{k'}}\left( {x,y} \right)dxdy}$, $\Omega _{kk'}^\mathcal{S}=\int_{\left( {x,y} \right) \in \mathcal{S}} {g_k^ * \left( {x,y} \right){e^{ - j\Delta {d_{k'}}\kappa \cos {\phi _{k'}}}}dxdy}$, and $D_{kk'}^\mathcal{S}=\iint_{\left( {x,y} \right) \in \mathcal{S}} {g_k^ * \left( {x,y} \right){g_{k'}}\left( {x,y} \right)dxdy}$, respectively.

\begin{Proposition}
In a C-LIS, the achievable SE of the $k$th user for a Ricean fading scenario can be approximated as
\begin{equation}\label{R_app}
{\mathsf{R}_k^{\rm{r}}} \approx {\log _2}\left( {1 + \frac{{{p_k}{\rm{P}}{{\rm{L}}_k}\pi {R^2}\left( {\pi {R^2} + {{ {{\left({1 + \gamma_k }\right)^{-2}}} }}} \right)}}{{\pi {R^2}{\sigma ^2} + \sum\nolimits_{k' \ne k} {{p_{k'}}{\rm{P}}{{\rm{L}}_{k'}}{{\rm{I}}_{kk'}}} }}} \right),
\end{equation}
where ${\rm I}_{kk'}$ is defined in (\ref{A_Eff_ricain}) at the top of next page with $\iota_k = \frac{{\sqrt {x_k^2{\text{ + }}y_k^2} }}{{d_k^c}}$.
\begin{figure*}[!t]
\vspace{-0.2cm}
\begin{align}\label{A_Eff_ricain}
{\rm I}_{kk'}&\triangleq \frac{1}{{\left( {1 + {\gamma _k}} \right)\left( {1 + {\gamma _{k'}}} \right)}}\notag \\
&\times\left( {{\gamma _k}{\gamma _{k'}}{\text{B}}{{\left( {R,\kappa ,{\chi _{kk'}}} \right)}^2} \!+\! {\gamma _k}{\text{B}}{{\left( {R,\kappa ,{\iota _k}} \right)}^2}\! + \!{\gamma _{k'}}{\text{B}}{{\left( {R,\kappa ,{\iota _{k'}}} \right)}^2}}\! +\!{\sqrt {{\gamma _k}{\gamma _{k'}}} {\pi ^2}{R^2}{\text{B}}\left( {R,\kappa ,{\chi _{kk'}}} \right)\! +\! \frac{1}{4}{\pi ^2}{R^2}\left( {{\pi ^2}{R^2} + 1} \right)} \right)
\end{align}
\vspace{-0.2cm}
\hrulefill
\end{figure*}
The overall sum SE across $K$ users then approximates
\begin{equation}\label{R_k_sum}
{\mathsf{R}^{\text{C-LIS}}_{{\text{total}}}} \approx \sum\nolimits_{k = 1}^K {\mathsf{R}_k^{\rm{r}}}.
\end{equation}
\end{Proposition}
\begin{IEEEproof}
We start by calculating $\Omega _{kk'}^\mathcal{S}$. Noting that $g_{k'}(x,y)$ is independent with ${e^{j\Delta {d_k}\kappa \cos {\phi _k}}}$ at every point on the surface, the value of $\Omega _{kk'}^\mathcal{S}$ is not constant but follows normal distribution as
$\Omega _{kk'}^\mathcal{S} \sim \mathcal{CN}\Big(0, {\big( {\int_{\left( {x,y} \right) \in \mathcal{S}} {{e^{j\Delta {d_k}\kappa \cos {\phi _k}}}dxdy}} \big)^2}\Big)$. Utilizing the same method as in (\ref{B_kk_integ}), we can obtain $\Omega _{kk'}^\mathcal{S} \sim \mathcal{CN}\Big(0, {\text{B}}{{\left( {R,\kappa ,{\iota _k}} \right)}^2} \Big)$ and $\Omega _{kk'}^\mathcal{S} \sim \mathcal{CN}\Big(0, {\text{B}}{{\left( {R,\kappa ,{\iota _{k'}}} \right)}^2} \Big)$.
For $D_{kk'}^{\mathcal{S}}$, when $k'=k$, we have $D_{kk}^{\mathcal{S}}\sim \mathsf{Gamma}(\pi R^2,1)$, and when $k'\neq k$, the value of $\left|g_k^ * \left( {x,y} \right){g_{k'}}\left( {x,y} \right)\right|$ follows a \emph{complex-valued central-normal distributions} with mean and variance being $\pi/2$ and $\pi/4$, respectively \cite{Wells1962}. Therefore, we finally have
$(\mathsf{E}\left[ {D_{ij}^\mathcal{S}} \right],\mathsf{var}\left[ {D_{ij}^\mathcal{S}} \right])=({\pi ^2}{R^2}/{2},{\pi ^2}{R^2}/{4})$ for $i\ne j$, and $(\mathsf{E}\left[ {D_{ij}^\mathcal{S}} \right],\mathsf{var}\left[ {D_{ij}^\mathcal{S}} \right])=(\pi {R^2},{\pi {R^2}})$ for $i= j$.
Replacing $\Sigma_{kk'}^{\mathcal{S}}$ in (\ref{Rate_general}) by $\tilde{\Sigma}_{kk'}^{\mathcal{S}}$, substituting expectation results into it and after some manipulations, we obtain the result.
\end{IEEEproof}

The approximation in (\ref{R_app}) becomes
more accurate with increasing surface area \cite[Lemma 1]{Qizhang2014}. Thus, in C-LIS systems, due to the large surface area, this approximation will be particularly accurate.}

%It is worth noting that, due to the intrinsic property of $J_1{\cdot}$, $\rm{B}(R,\kamma,\iota_{k})$ approximates to 0 when $\iota_{k}\\rightarrow 1$, which indicates that the interference from the NLoS path is almost non-existent if distance

{Fig. 4 verifies our theoretical analysis, where the curve $\lambda = 0_+$ represents the upper bound in (\ref{R_k_sum_inf}). {We assume the same transmit power and same Ricean factor across $10$ users, and denote $\rho = \left\{\frac{p_k}{\sigma^2}\right\}_{k=1,\ldots,K}$}.\footnote{The parameter $\rho$ refers to the ratio of the signal power at the transmitter to the noise power at LIS. The reason for using $\rho$ instead of conventional SNR is that, with randomly deployed users, the SNR at LIS is unknown due to the effect of the path loss.}
From Fig. 4a, we observe a significant degradation of the SE from pure LoS channel to Ricean channel, which indicates the great impact of interference from the NLoS path.
Moreover, it can be seen that the achievable sum SE is closer to the upper bound with a higher frequency band or with a larger LIS surface for both pure LoS channel and Ricean channel.
An important observation is that, when $R$ increases to 500\,m, for the case that $\lambda = 0.2$\,m (i.e., 1.5\,GHz), a small gap to the upper bound can be observed for the LoS channel while there is nearly no gap for the Ricean channel. This showcases that the interference is dominated by the NLoS path at high frequency bands.
Besides, as there is hardly any gap between $\lambda=0.005$\,m and the upper bound, the expression in (\ref{R_k_sum_inf}) can perfectly approximate the performance for mmWave frequencies in LoS scenarios, even with finite LIS.
Fig. 4b compares the per-user achievable SE against $\lambda$ for $K=10$ and $K= 20$ scenarios.
We can observe that the per-user achievable SE can reach the upper bounds in any scenarios when the wavelength is sufficient short.
In addition, the per-user SEs for $10$\,users and $20$\,users converge to the same value in pure LoS scenarios, while a small gap can be observed in Ricean fading cases.
More importantly, the gap between the SE in LoS and in Ricean fading increases remarkably when increasing the power from $\rho=60$\,dB to $\rho=90$\,dB.
This is due to the fact that the interference from the LoS path can be nearly cancelled at high frequency band, so that the SE in pure LoS scenarios grows linearly with increasing transmit power.}

\subsection{Sum SE Maximization}

As the SE performance of each user is coupled together due to ${\rm{\tilde B}}{{\left( {R,\kappa ,{\chi _{,kk'}}} \right)}}$, and with the fact that ${\rm{\tilde B}}{{\left( {R,\kappa ,{\chi _{kk'}}} \right)}}$ is not convex, a possible way to solve the sum SE (as given by (\ref{R_k_sum})) maximization problem is the following
\begin{subequations}\label{max_C_rate_all}
\begin{align}
\label{R_sum_1}\mathop{\text{maximize}}\limits_{\kappa,\vartheta } \;\;&{{\mathsf{R}}^{\text{C-LIS}}_{\text{total}}} \hfill\\
\label{con_1}{\text{s}}{\text{.t}}{\text{.}}\;\;&\kappa  \in \left[ {{\kappa _{\min }},{\kappa _{\max }}} \right],\;\;\vartheta \in \left[-\pi,\pi\right].
\end{align}
\end{subequations}
is brute-force searching, where ${\kappa _{\min }}$ and ${\kappa _{\max}}$ are the minimum and maximum values. Since we need $\frac{K(K-1)}{2}$ operations to evaluate all the ${\rm{\tilde B}}{{\left( {R,\kappa ,{\chi ^\vartheta_{kk'}}} \right)}}$ with fixed $\kappa$ and $\vartheta$, the overall computational complexity is $\mathcal{O}\left(\frac{ILK^2(K-1)}{2}\right)$, where $I=\frac{{\kappa_{\max }}-{\kappa_{\min}}}{\Delta \kappa}$ and $L=\frac{2\pi}{\Delta \vartheta}$ with searching step ${\Delta \kappa}$ and ${\Delta \vartheta}$.
Fortunately, based on the result in \emph{Corollary} 1, in general, the achievable sum SE grows monotonically by increasing the frequency band. Therefore, the complexity can be remarkably reduced by setting $\kappa = \kappa_{\min}$, and the searching complexity for the maximization problem
\begin{subequations}\label{max_C_rate}
\begin{align}
\label{R_sum_theta}\mathop{\text{maximize}}\limits_{\kappa_{\min},\vartheta } \;\;&{{\mathsf{R}}^{\text{C-LIS}}_{\text{total}}} \hfill\\
\label{con_theta}{\text{s}}{\text{.t}}{\text{.}}\;\;&\vartheta \in \left[-\pi,\pi\right],
\end{align}
\end{subequations}
becomes only $\mathcal{O}\left(\frac{LK^2(K-1)}{2}\right)$.
%More importantly, although the sum rate maximizing in (\ref{max_C_rate_all}) refers that there is a optimal $\lambda$ that minimizes the interference, we find that the achievable rate almost increase monotonously with respect to $R$ and $\lambda$.

\section{Performance Analysis of D-LIS}

%Clearly, to deploy such a LIS with hundreds of square metre is not engineering implementable, we hence consider a distributed layout, in which the $M$ LIS is split into $L (L>K)$ units, and each LIS unit only serves a particular user. The advantage of the distributed LIS is that, for each unit, it is allowed to intelligently control its angle to minimize the interference according to user's position.

To deploy a centralized LIS spanning tens or even hundreds of square meters is not always feasible; for this reason, we hereafter consider a distributed topology, in which $M (M>K)$ same-size LIS-units are randomly spread over a large area and are connected to a centralized baseband unit.\footnote{The optimal pattern to deploy the D-LIS units remains an interesting and open problem. A rule of thumb is to deploy more D-LIS units in places where it is more likely to have more users. This can be realized in practice as long as the user pattern can be observed by collecting statistical information ahead of the network design.}
In D-LIS systems, each LIS-unit serves a particular user.
The advantage of the distributed LIS is twofold: (a) as each LIS-unit is assigned to one particular user, it is possible control its orientation to minimize the interference for the target user without considering the impact on others; hence, the orientation control complexity can be largely reduced; (b) the distributed system can enhance the signal strength by reducing the propagation distance, and through meticulous user association and power control, the signal quality can be further improved.
In this section, we study the achievable SE for a D-LIS, and we aim to design effective schemes to maximize the total sum SE and maximize the minimum user SE by proposing new user association, orientation control and power control schemes.

We first evaluate the SE of the $k$th user achieved at arbitrary LIS-unit, e.g., the $m$th LIS-unit, with respect to LIS orientation based on the result in \emph{Proposition} 3.

\begin{Proposition}
Denote by $\vartheta_{m}$ the adjust angle of the ${m}$th LIS-unit. The achievable SE of the $k$th user at the ${m}$th LIS-unit equals
\begin{align}\label{Dis_per_rate}
&{\mathsf R}_{k,m} \notag\\
&\!= \!{\log _2}\!\left(\! {1\! + \! \frac{{{p_k}{\text{P}}{{\text{L}}_{k,m}}}}{{\frac{\sigma ^2}{{\pi R_{\rm{d}}^2}}\! +\! \sum\limits_{k' \ne k} {{p_{k'}}{\text{P}}{{\text{L}}_{k',m}}{\rm{\tilde B}}{{\left( \!{{R_{\rm{d}}},\kappa ,\chi _{m,kk'}^{{\vartheta_{m}}}}\! \right)}^2}} }}} \!\right),
\end{align}
where $R_{\rm{d}}$ is the radius of the LIS-unit in D-LIS, ${\text{P}}{{\text{L}}_{i,j}}$ is the path loss from the $i$th user to the $j$th LIS-unit, and $\chi _{m,kk'}^{{\vartheta_{m}}}$ has the same form as in \emph{Proposition} 3 by simply using the coordinates of the $m$th LIS-unit.
%Therefore, the overall sum SE can be obtained as
%\begin{equation}\label{Dis_sum_rate}
%{{\mathsf R}^{\text{D-LIS}}_{{\text{total}}}} = \sum\nolimits_{k = 1}^K {{\mathsf R}_{k,m_k}}.
%\end{equation}
\end{Proposition}
\begin{IEEEproof}
The result can be directly obtained by substituting \emph{Definition} 1 and \emph{Proposition} 3 into (\ref{Rate_general}).
\end{IEEEproof}

Different from the expression in (\ref{R_k}) in which the achievable SEs of each user couple together due to ${\rm{\tilde B}}\left( {R,\kappa ,\chi _{kk'}^\vartheta } \right)$, the per-user achievable SE at each LIS-unit is independent across $M$ LIS-units, which allow us to adjust each unit separately.

\subsection{User Association}
%Besides the centralized and coordinated LIS, distributed layout is consider in this paper, in which $M (M>K)$ same-size LIS spread over a large area forming a large distributed LIS system, and serve $K$ users simultaneously. We assume each LIS only serve a particular user, hence, the aim is to find $K$ best LIS that maximize the sum-rate and minimum user-rate, respectively.

By harnessing the result in \emph{Proposition} 5, we elaborate on user association for sum SE maximization and minimum user SE maximization.

Denote by $\bf S$ the $K\times M$ LIS selection matrix, whose $(k,m)$th element is $s_{k,m}\subset[0,1]$ with $s_{k,m} =1$ representing that the $k$th user is associated to the $m$th LIS, and $s_{k,m} =0$ otherwise.
%According to \emph{Proposition} 4, the rate of the $k$th user achieved with the $m$th LIS is given as
%\begin{equation}\label{dis_LIS_mk}
%R_k^m = {\log _2}\left( {1 + \frac{{{p_k}{\text{P}}{{\text{L}}_{k,m}}}}{{\frac{1}{{\pi {R^2}}}{\sigma ^2} + \sum\limits_{k' \ne k} {{p_{k'}}{\text{P}}{{\text{L}}_{k',m}}{{{\rm{\tilde B}}}_m}{{\left( {R,\kappa ,{\chi _{{\text{Los}},kk'}}} \right)}^2}} }}} \right),
%\end{equation}
%where ${\text{P}}{{\text{L}}_{k',m}}$ is the pathloss between the $k'$th user and the $m$th LIS, ${{{\rm{\tilde B}}}_m}{{\left( {R,\kappa ,{\chi _{{\text{Los}},kk'}}} \right)}^2}$ is the normalized LIS response at the $m$th LIS.
Given the achievable SE expression in (\ref{Dis_per_rate}), we formulate the sum SE maximization problem as
\vspace{-0.1cm}
\begin{subequations}\label{SR_Ori}
\begin{align}
\label{ob_ori}\mathop {{\text{maximize}}}\limits_{\mathbf{S}}\;\; &\sum\nolimits_k\sum\nolimits_m {{s_{k,m}}\mathsf{R}_{k,m}} \\
\label{select_ele}{\rm{s}}{\rm{.t}}{\rm{.}}\;\;&{s_{k,m}} \subset \left[ {0,1} \right],\;\;\;\;\;\;\forall k,\;m,\\
\label{select_row}&\left\| {\bf s}_k \right\|_0  = 1,\;\;\;\;\;\;\;\;\;\forall k,\\
\label{select_column}&\left\| \left\{{\bf S}\right\}_m \right\|_0 \subset [0,1], \;\;\forall m,
\vspace{-0.1cm}
\end{align}
\end{subequations}
where ${\bf s}_k$ and $\left\{{\bf S}\right\}_m$ represent the $k$th row and the $m$th column of $\bf S$, respectively. The constraint (\ref{select_row}) ensures that each user is served by a LIS, and (\ref{select_column}) guarantees that each LIS serves no more than one user. Similarly, the minimum user SE maximization problem can be written as
\vspace{-0.1cm}
\begin{subequations}\label{MMax_Ori}
\begin{align}
\label{ob_ori_mmax}\mathop {{\text{maximize}}}\limits_{\mathbf{S}}\;\;& {\mathop {\min }\limits_{k,m} \left\{ {{s_{k,m}}\mathsf{R}_{k,m}} \right\}}\\
\label{select_same}{\rm{s}}{\rm{.t}}{\rm{.}}\;\;\;&(\ref{select_ele}),(\ref{select_row}),(\ref{select_column}).
\vspace{-0.1cm}
\end{align}
\end{subequations}

Note that, both the sum SE maximization problem and minimum user SE maximization problem are nonconvex even without the discrete constraints, whose optimal result can only be solved via searching. When $M$ and $K$ are large values, the computational complexity is unaffordable.
Therefore, we now propose a suboptimal iterative user association algorithm to reduce this complexity.

By noting that the interference can be largely reduced by adjusting the orientation of the LIS, i.e., ${\rm{\tilde B}}{{\left( {R_{\rm{d}},\kappa ,{\chi ^{\vartheta_m}_{m,kk'}}} \right)}}$ can be ignored when $k \neq k'$, a LSF-based user association (LUA) scheme is then proposed. When no interference is considered, the maximization problems in (\ref{SR_Ori}) and (\ref{MMax_Ori}) can be rewritten in the following forms
\vspace{-0.1cm}
\begin{align}\label{SR_var}
\mathop {{\text{maximize}}}\limits_{\mathbf{S}}\;\; &\sum\nolimits_k\sum\nolimits_m {{s_{k,m}}{\text{P}}{{\text{L}}_{k,m}}}\\
{\rm{s}}{\rm{.t}}{\rm{.}}\;\;&(\ref{select_same}),\notag
\vspace{-0.1cm}
\end{align}
and
\vspace{-0.1cm}
\begin{align}\label{R_mmax_var}\mathop {{\text{maximize}}}\limits_{\mathbf{S}} \;\;&{\mathop {\min }\limits_{k,m} \left\{ {{s_{k,m}}{\text{P}}{{\text{L}}_{k,m}}} \right\}} \\
{\rm{s}}{\rm{.t}}{\rm{.}}\;\;&(\ref{select_same}),\notag
\vspace{-0.1cm}
\end{align}
respectively. Even when adopting the LUA, the problems are still not solvable since the constraints in (\ref{select_same}) are discrete, which makes the optimization problems non-convex. Hence, we design a reweighted $\ell_1$-norm iterative method to approximate the constraint \cite{Cands2008}. As the $s_{k,m}$ is either 0 or 1, we hence approximate the coefficient in the following form
\vspace{-0.1cm}
\begin{equation}\label{app_s}
\left\|{s}_{k,m}\right\|_0 \approx \left\|\omega_{k,m}{\tilde s}_{k,m}\right\|_1,
\vspace{-0.1cm}
\end{equation}
where ${\tilde s}_{k,m} \in [0,1]$ is a continuous value, and $\omega_{k,m}=\frac{1}{{\tilde s}_{k,m}+\varrho}$ denotes the weight coefficient associated with ${\tilde s}_{k,m}$, in which $\varrho$ is a very small positive value that provides stability. It is straightforward to see that the right hand of (\ref{app_s}) will force the expression converge to either $0$ or $1$. Utilizing this approximation, the problems (\ref{SR_var}) and (\ref{R_mmax_var}) can, thus, be transformed as
\begin{subequations}\label{SR_var_var}
\begin{align}
\label{ob_var}\mathop {{\text{maximize}}}\limits_{\tilde{\mathbf{S}}}\;\; &\sum\nolimits_k\sum\nolimits_m {\omega_{k,m}{\tilde s}_{k,m}{\text{P}}{{\text{L}}_{k,m}}} \\
\label{select_ele_var}{\rm{s}}{\rm{.t}}{\rm{.}}\;\;&{{\tilde s}_{k,m}} \in \left[ {0,1} \right],\;\;\;\;\;\;\;\;\;\;\forall k,\;m,\\
\label{select_row_var}&\left\|{\boldsymbol\omega}_{k}\cdot{\tilde {\bf s}}_{k} \right\|_1  = 1,\;\;\;\;\;\;\;\;\;\;\;\forall k,\\
\label{select_column_var}&\left\| \left\{{\bf{\Omega}\cdot \bf S}\right\}_m \right\|_1 \in [0,1], \;\;\forall m,
\end{align}
\end{subequations}
and
\begin{subequations}\label{R_mmax_var_var}
\begin{align}
\label{ob_var_mmax}\mathop {{\text{maximize}}}\limits_{\tilde{\mathbf{S}}}\;\;&{\mathop {\min }\limits_{k,m} \left\{ {{{\omega}_{k,m}{\tilde s}_{k,m}}{\text{P}}{{\text{L}}_{k,m}}} \right\}} \\
\label{select_same_var}{\rm{s}}{\rm{.t}}{\rm{.}}\;\;&(\ref{select_ele_var}),(\ref{select_row_var}),(\ref{select_column_var}),
\end{align}
\end{subequations}
where the operator $\cdot$ represents the dot product, ${\bf{\Omega}} \in \mathbb{R}^{K\times M}$ is the weight matrix whose $(k,m)$th element is $\omega_{k,m}$, and ${\boldsymbol\omega}_{k}$ represents the $k$th column of $\bf{\Omega}$. Note that, with this approximation, the optimization problems (\ref{SR_var_var}) and (\ref{R_mmax_var_var}) are convex with fixed $\bf \Omega$, hence, we can develop an iterative method to achieve suboptimal LIS selection, where $\omega_{k,m}$ in each iteration is updated via the solution of ${\tilde s}_{k,m}$ from the previous iteration. According to the complexity analysis in \cite{gass2003linear}, the arithmetic complexity per iteration of our algorithm is $\mathcal{O}(K^{3.5})$. If we set the limit of the iterations as $N$, the overall complexity is then upper bounded by $\mathcal{O}(NK^{3.5})$. The procedure of LUA is detailed in \textbf{Algorithm} 1.

\newcommand{\algorithmicinitial}{\textbf{Initialization:}}
\newcommand{\algorithmicstep}{\textbf{Step}}
\newcommand{\STEP}{\item[\algorithmicstep]}
\newcommand{\INITIAL}{\item[\algorithmicinitial]}

\begin{algorithm}[!t]\label{algorithm1}
\caption{}
\begin{algorithmic}
{\INITIAL Weight matrix ${\bf \Omega} = {\bf 1}^{K\times M}$, iteration count $Count = 1$, maximum iteration number $N$, threshold $\tau$ and the parameter $\varrho$

\WHILE{$Count<=N$}
\STATE Solve problems (\ref{SR_var_var}) or (\ref{R_mmax_var_var}).
\STATE Update $\omega_{k,m}$ via $\omega_{k,m}=\frac{1}{{\tilde s}_{k,m}+\varrho}$,
\STATE $Count=Count +1$
\ENDWHILE
\STATE Set ${\tilde s}_{k,m}=1$ if ${\tilde s}_{k,m}\geq\tau$, $\forall k,m$; otherwise, ${\tilde s}_{k,m}=0$.}
\end{algorithmic}
\end{algorithm}

\subsection{Orientation Control}

Similar to C-LIS systems, the orientation of the LIS-unit has an important impact on the per-user SE user SE for D-LIS. However, a fundamental difference between C-LIS and D-LIS is that, by assigning each user to different LIS-unit, the per-user SEs are decoupled at $K$ LIS-units in D-LIS. This allows each LIS-unit to adjust its orientation to maximize the achievable SE for the user associated to it.

%Denote by $\dot{\mathbf{S}}$ the association result solved in problem (\ref{SR_var_var}) or (\ref{R_mmax_var_var}) previously,

We denote by $\mathcal{P}$ the $K$ pairs of association results solved in problem (\ref{SR_var_var}) or (\ref{R_mmax_var_var}) previously, in which the $k$th pair of user and LIS-unit is defined as ${\mathcal{P}_k}:\{k,m_k\}$.
%and assume the $k$th is associated to the $m_k$th LIS-unit.
Therefore, the OC can be split into $K$ subproblems, in which each subproblem is formulated as
\begin{subequations}\label{max_rate_varphi}
\begin{align}
\label{R_sum_var}\mathop {\text{maximize}}\limits_{{\vartheta _{m_k}}} \;\;&\mathsf{R}_{\mathcal{P}_k}  \hfill\\
\label{con_phi}{\text{s}}{\text{.t}}{\text{.}}\;\;&{\vartheta _{m_k}} \in \left[ \pi, \pi \right].
\end{align}
\end{subequations}
%where $\dot{s}_{k,m}$ is the $(k,m)$th element of $\dot{\mathbf{S}}$ which is a constant value equaling either 0 or 1.
%As shown in (\ref{Dis_per_rate}), the problem in (\ref{R_sum_var}) is equivalent to a classic weighted sum problem
%\begin{equation}\label{equivalent_rate}
%\mathop {{\text{min}}}\limits_{{\vartheta _k}} \;\sum\limits_{k' = k} {{p_{k'}}{\text{P}}{{\text{L}}_{k'}}{\rm{\tilde B}}{{\left( {{R_k},\kappa ,\chi _{{\text{Los}},kk'}^{{\vartheta _k}}} \right)}^2}}.
%\end{equation}
where $\mathsf{R}_{\mathcal{P}_k}$ refers to expression in (\ref{Dis_per_rate}), and ${\vartheta _{m_k}}$ is the adjust angle of the $m_k$th LIS-unit. However, even with the fact that the per-user SE across users are decoupled at the LIS side, the objective function (\ref{R_sum_var}) is still non-convex due to
the Bessel function. Therefore, the brute-force searching is required for the solving problem, whose complexity is ${\mathcal O}(LK)$, and the overall computational complexity to solve $K$ subproblems is ${\mathcal O}(LK^2)$.

Instead of brute-force searching, a suboptimal algorithm is proposed to reduce the complexity. By noting that the interference is mainly caused by the nearest user, utilizing the closed-form results in \emph{Property} 3, we are able to reduce the overall interference by simply minimizing this interference. The algorithm is then detailed in \textbf{Algorithm 2}, in which each LIS-unit adjusts their orientation to minimize the interference caused by the user who has the minimum ${\chi ^{\vartheta_{m_k}}_{{m_k},kk'}}$ with their associated user.
As the brute-searching over orientation angle has been substantially decreased, the complexity of the proposed algorithm is ${\mathcal O}(K^2)$.

\begin{algorithm}[!b]\label{algorithm2}
\caption{}
\begin{algorithmic}
\INITIAL $\mathcal{P}$, $\kappa$, $R_{\rm{d}}$.
\FOR{$k=1$ to $K$}
\STATE Calculate ${\chi ^{\vartheta_{m_k}}_{{m_k},kk'}}$ for $k'\neq k, k'\in K$, and select smallest ${\chi ^{\vartheta_{m_k}}_{{m_k},kk'}}$ as target
\STATE Calculate corresponding $\left|{\xi ^{\vartheta_{m_k}}_{{m_k},kk'}}\right|$ and $\left|\varpi^{\vartheta_{m_k}}_{{m_k},kk'}\right|$
\IF {$j_{1,n}$ exists in the range of $\left[{R}\kappa | {{\xi ^{\vartheta_{m_k}}_{{m_k},kk'}}} |,{R}\kappa |{\varpi ^{\vartheta_{m_k}}_{{m_k},kk'}}|\right]$}
\STATE Set $\vartheta_{m_k}$ according to $\vartheta = \tfrac{1}{2}\arctan {\bar v}$;
\ELSE
\STATE find $\min\left\{{\rm{\tilde B}}\left( {{R_{\rm{d}}},\kappa ,\left|{\xi ^{\vartheta_{m_k}}_{{m_k},kk'}}\right|} \right),{\rm{\tilde B}}\left( {{R_{\rm{d}}},\kappa ,\left|\varpi^{\vartheta_{m_k}}_{{m_k},kk'}\right|} \right)\right\}$, and set $\vartheta_{m_k}$ accordingly via (\ref{var_theta_solu}).
\ENDIF
\ENDFOR
\end{algorithmic}
\end{algorithm}

\subsection{Max-Min Power Control}

The LSF-based user association harnesses the advantages of D-LIS in exploiting the multiuser diversity, while PC can further enhance the system performance by providing uniform coverage.

We consider that the PC procedure is performed immediately after OC, in which the knowledge of LIS assignment is already known at the central baseband unit. Therefore, we denote by ${\dot{\vartheta} _{m_k}}$ the orientation control result at the ${m_k}$th LIS-unit; the max-min PC problem can, thus, be formulated as
\begin{subequations}\label{Power_control}
\begin{align}
\label{ob_max_min}\mathop {{\text{maximize}}}\limits_{{\tau_k}}\;\;&{\mathop {\min }\limits_{k, \forall k} \mathsf{SINR}_k} \\
\label{weight_pc}{\rm{s}}{\rm{.t}}{\rm{.}}\;\;& 0\leqslant\tau_k\leqslant1,\;\;\; k= 1,\ldots,K,
\end{align}
\end{subequations}
where
\begin{align}
\mathsf{SINR}_k\!=\!{\frac{{{\tau _k}{p_k}{\text{P}}{{\text{L}}_{\mathcal{P}_k}}}}{{\frac{\sigma ^2}{{\pi R_{\rm{d}}^2}}\!+\! \sum\limits_{k' \ne k}\! {{\tau _{k'}}{p_{k'}}{\text{P}}{{\text{L}}_{k',m_k}}{\rm{\tilde B}}{{\left( \!{{R_{\rm{d}}},\kappa ,\chi _{{m_k},kk'}^{{\dot{\vartheta} _{m_k}}}} \!\right)}^2}} }}}
\end{align}
with $\tau_k$ being the power control coefficient of the $k$th user.

We solve the power allocation problem for a given LIS selection matrix and orientation phases which can
be casted as a max-min SINR maximization problem.
%{\rl Note that the minimum user SE maximization is equivalent to the minimum SINR maximization due to the positive correlation between SE and SINR.}
Without loss of generality, by introducing an additional factor $t$, the problem in (\ref{Power_control}) can be reformulated as
\begin{subequations}\label{Re_Power_control}
\begin{align}
\label{re_ob_max_min}
\mathop {{\text{maximize}}}\limits_{{\tau _k},\,t} \;\;&t \\
\label{re_weight_pc}{\rm{s}}{\rm{.t}}{\rm{.}}\;\;&t\leqslant\mathsf{SINR}_k,\;(\ref{weight_pc}).
\end{align}
\end{subequations}
It is straightforward that, for a given $t$, all the inequalities in constraint (\ref{re_weight_pc}) are linear, making the problem (\ref{Re_Power_control}) a quasi-linear problem \cite{7827017}. Hence, such problem can be efficiently solved by using the bisection method and solving a sequence of linear feasibility problems in (\ref{re_weight_pc}). The algorithm is detailed in \textbf{Algorithm 3}. Note that, given a tolerance $\epsilon$, it takes $V= \log_{\epsilon}\left(t_{\max}-t_{\min}\right)$ iterations for parameter $t$ to converge, and the complexity of $\mathcal{O}(K^3)$ for \emph{Gaussian elimination} to solve the equation sets in each iteration. Thus, the overall complexity is equivalent as $\mathcal{O}(VK^3)$.

Based on the complexity analysis of user association, orientation control and power control, we note that the overall complexity among these three procedures is dominated by the user association, which underlines that the goal of evaluating the sum SE maximization or minimum SE maximization has a complexity of $\mathcal{O}(NK^{3.5})$.

\begin{algorithm}[!t]\label{algorithm3}
\caption{}
\begin{algorithmic}
\INITIAL choose the initial values of $t_{\max}$ and $t_{\min}$, input the LIS selection matrix $\dot{\mathbf{S}}$, set a tolerance value $\epsilon>0$.

\WHILE{$t_{\max}-t_{\min}>\epsilon$}
\STATE 1) $t=\frac{t_{\max}-t_{\min}}{2}$ and solve the following a sequence of equations:
\[\mathsf{SINR}_k= t,\;\;k=1,\ldots,K.\]
%\IF {$0\leqslant \tau_k\leqslant 1,\;\forall k$}
%\STATE $t_{\min} = t$
%\ELSE
%\STATE $t_{\max} = t$
%\ENDIF
\STATE 2) $t_{\min} = t$ if $0\leqslant \tau_k\leqslant 1,\;\forall k$, and $t_{\max} = t$, otherwise.
\ENDWHILE
\end{algorithmic}
\end{algorithm}

\section{Numerical Results}

In our simulations, we study the performance of C-LIS and D-LIS. We assume that, in C-LIS, a large LIS is located at the center of a square of size 1\,$\text{km}^2$, while $M$ LIS-units are randomly deployed in the same size of area in D-LIS. In the following, all the simulations are conducted for an operating frequency of 2 GHz.

\begin{figure*}[!t]
	\centering
	\includegraphics[width=16.5cm]{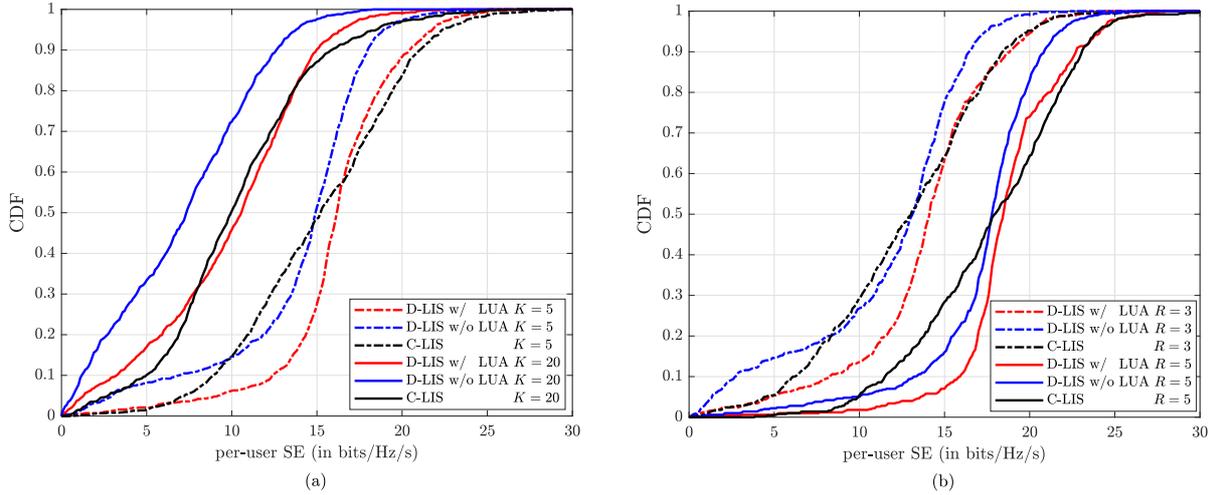}
	\caption{The CDFs of the achievable SE are evaluated for C-LIS and D-LIS, in which (a) compares the CDFs under different users with $R=5$, and (b) compares the CDFs under different surface areas with $K=5$. The results are shown for $M=20$.}
\end{figure*}

\begin{figure*}[!b]
	\centering \includegraphics[width=16.5cm]{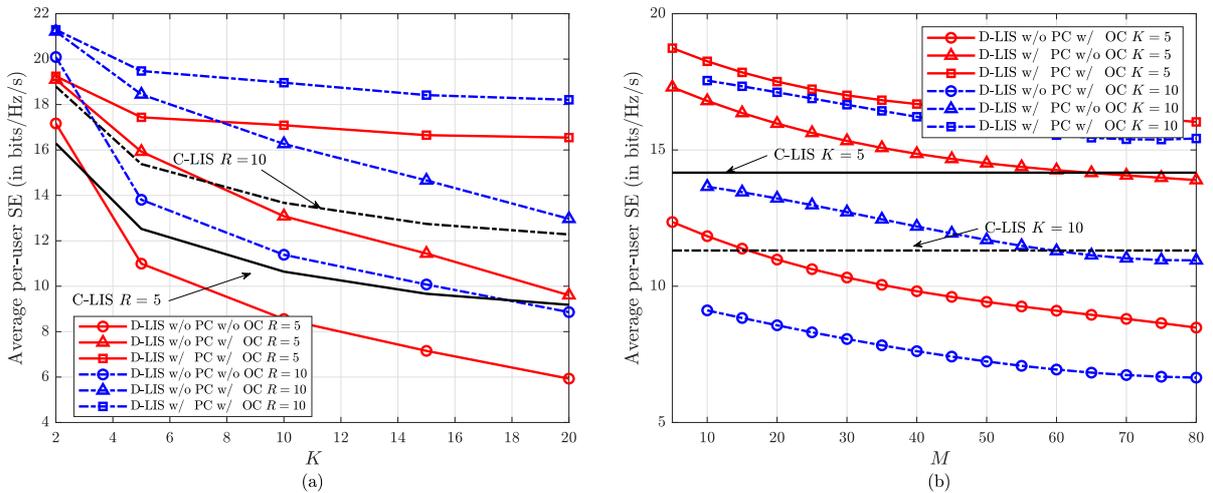}
	\caption{The average per-user SE evaluated with C-LIS and D-LIS, in which (a) compares the average per-user SE with respect to the number of users, and (b) compares the average per-user SE with respect to the number of LIS-units. The results are shown for $\rho = 110$\,dB, and averaged over 100 runs.}
\end{figure*}

To illustrate the importance of user association, we compare the CDF of achievable per-user SE in D-LIS and C-LIS for different scenarios in Fig. 5. In the simulations, there are 20\,LIS-units in the D-LIS system, and the overall surface area equals to the area of LIS in C-LIS for the sake of fairness.
The LUA aiming to maximize the sum SE is simulated in Fig. 5a, while the LUA used for maximizing the minimum SE is depicted in Fig. 5b.
From Fig. 5a, an obvious observation is that the proposed LUA algorithm is more effective in the scenarios with fewer users. For example, when $K=5$, the 95\%-likely achievable SE for D-LIS with LUA is about 10\,bits/Hz/s which is 8\,bits/Hz/s higher than for D-LIS without LUA, while such advantage reduces to 1\,bits/Hz/s for the 20\,user case. More importantly, we observe that C-LIS outperforms D-LIS for the 20\,user scenario while the opposite situation occurs with $K=5$.
%It is intuitively that C-LIS can provide higher quality service than D-LIS with fewer users, while the situation is opposite when serves 20\,users.
The reason is that, with more users, a higher spatial resolution LIS is required to distinguish the users who are closely located. Therefore, a D-LIS, whose LIS-units have a small area, is not the most appropriate solution for this particular scenario.
Fig. 5b shows the CDF of the achievable SE with different surface area for $K=5$.
We can see that by increasing the radius of surface from $3$\,m to $5$\,m, the 95\%-likely per-user SE can be improved over 200\%.
Besides, we clearly observe that, by applying LUA, D-LIS is superior to C-LIS for both $R=3$\,m and $R=5$\,m scenarios in terms of the per-user achievable SE, and such advantage is more significant with larger surface area.
%For instance, with RAS algorithm, the 95\%-likely D-LIS per-user rate is 4\,bits/Hz/s higher than C-LIS per-user rate when $R=10$\,m, while there appears no advantage for the $R=3$\,m cases.
%This result implies that, when only LUA is used, the distributed layout of LIS system is more suitable for the scenarios with sparse users

%
%\begin{figure}[!h]
%	\centering
%	\includegraphics[width=12.5cm]{Per_user_wrt_K.pdf}
%	\caption{The average per-user rate evaluated with C-LIS and D-LIS with respect to the number of users. The results are shown for $\rho = 110$\,dB, and averaged 100 runs.}
%\end{figure}
%
%\begin{figure}[!h]
%	\centering
%	\includegraphics[width=12.5cm]{Per_user_wrt_M.pdf}
%	\caption{The average per-user rate evaluated with C-LIS and D-LIS with respect to the number of LIS-units. The results are shown for $\rho = 110$\,dB, and averaged 100 runs.}
%\end{figure}

%

\begin{figure}[!t]
\centering
\includegraphics[width=8.7cm]{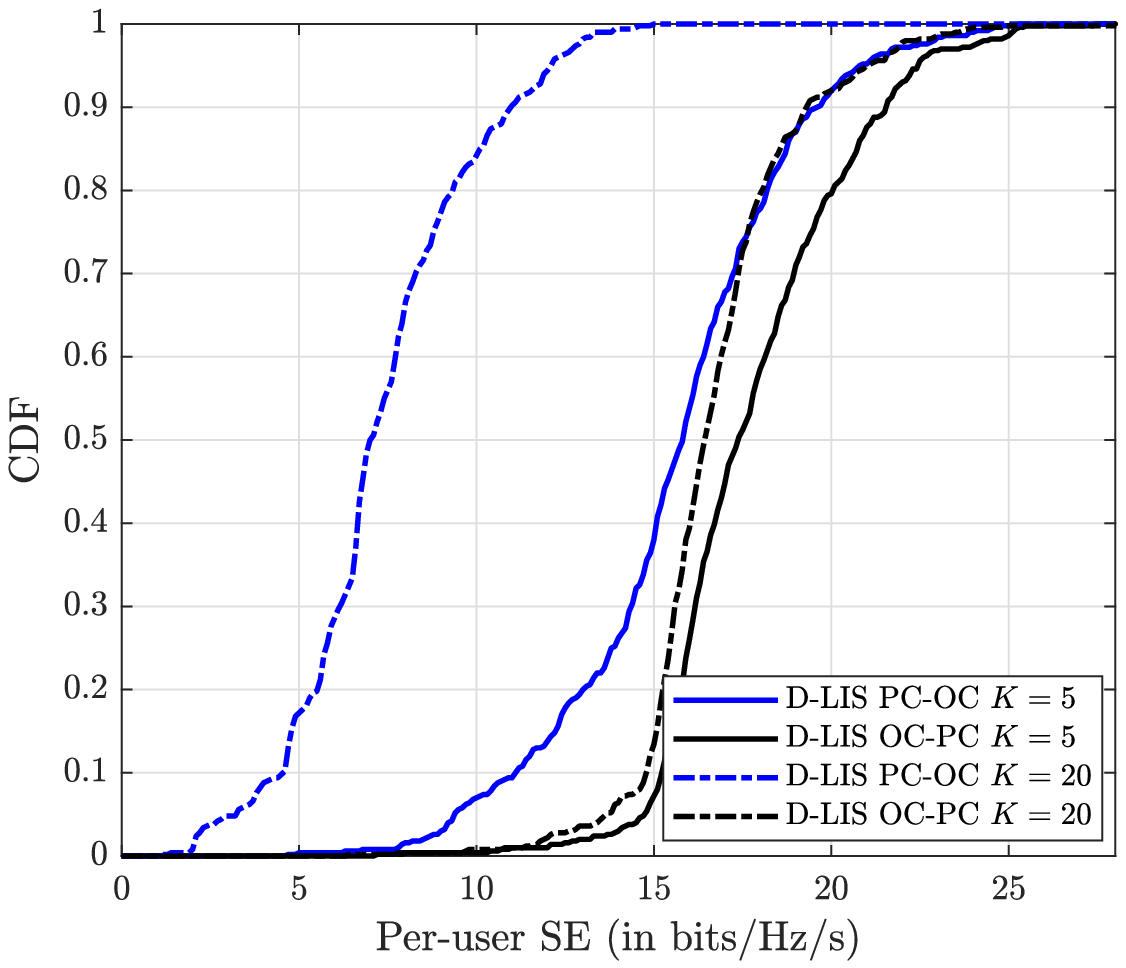}
\caption{The CDFs of the achievable SE are evaluated for D-LIS. The results are shown for $M=20$.}
\end{figure}

\begin{figure}[!t]
	\centering
	\includegraphics[width=9cm]{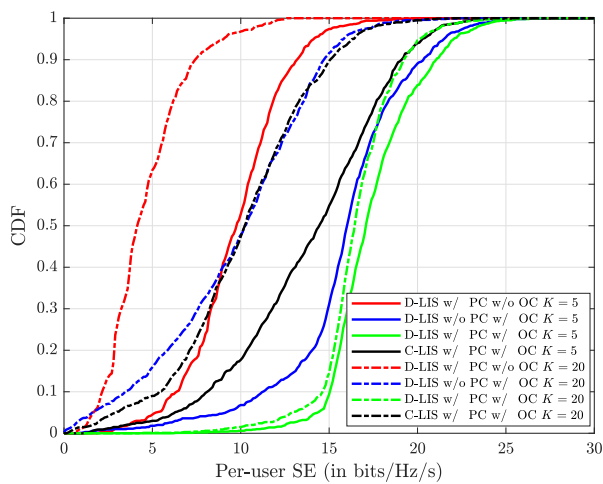}
	\caption{The CDFs of the achievable per-user SE are evaluated for C-LIS and D-LIS after LUA for maximizing the minimum user SE. The overall surface area of D-LIS equals to the area of C-LIS. The results are shown for 100 runs.}
\end{figure}

\begin{figure*}[!b]
\centering
\includegraphics[width=16.5cm]{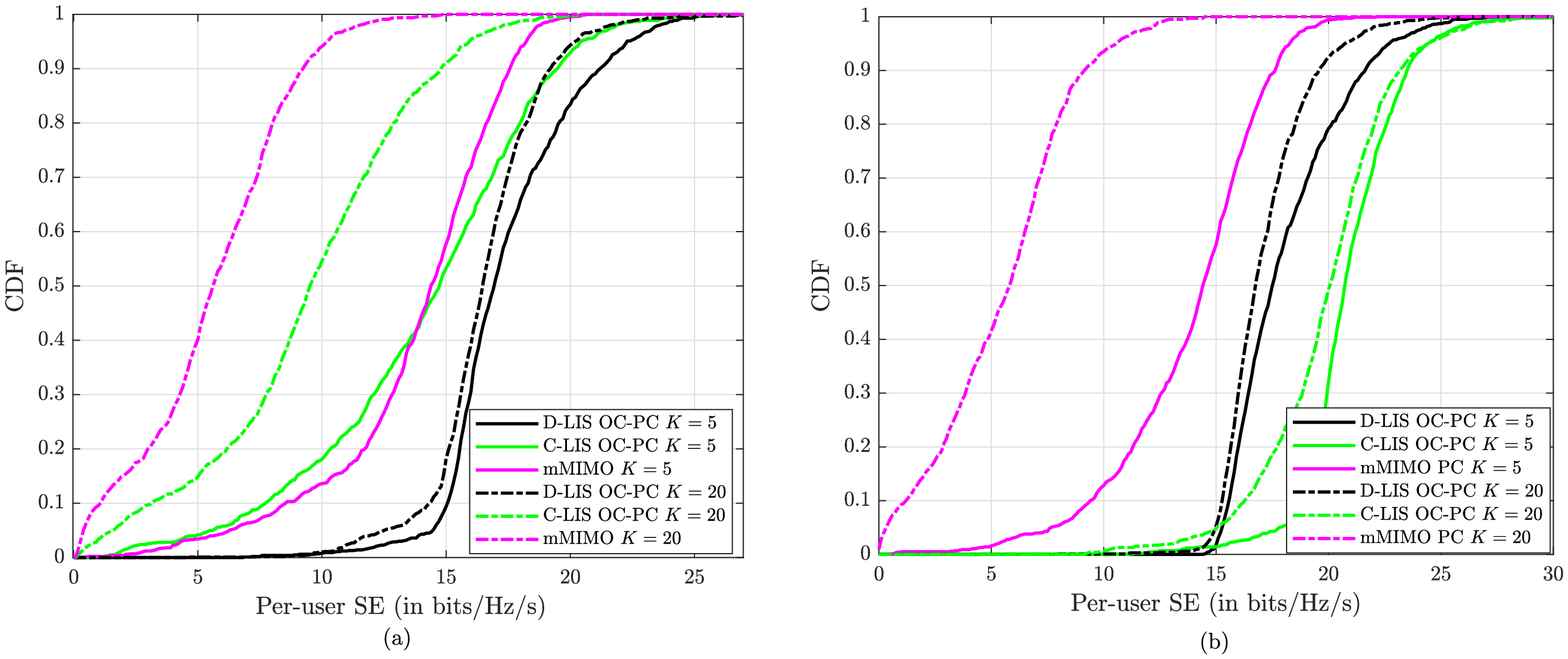}
\caption{The CDFs of the achievable SE are evaluated for C-LIS, D-LIS and mMIMO, in which (a) compares the CDFs in microwave band (2\,GHz), and (b) compares the CDFs in mmWave band (50\,GHz). The results are shown for $M=20$.}
\end{figure*}

We now illustrate the performance of the proposed max-min PC and OC algorithm.
Fig. 6a and Fig. 6b show the average per-user SE after LUA with respect to the number of users and the number of LIS-units, respectively.
Firstly, it can be observed from Fig. 6a that C-LIS outperforms D-LIS for most of scenarios if no resource allocation algorithms are applied, whereas the performance of D-LIS is significantly superior to C-LIS when OC and PC are considered.
In addition, Fig. 6a shows that the SE gain provided by PC and OC increases significantly with an increasing number of users.
For instance, with $R=5$\,m, only 2\,bits/Hz/s gain can be observed for the two-user scenario while such gain increases to about 11\,bits/Hz/s when 20\,users are served.
In order to identify the optimal number of LIS-units in D-LIS, we illustrate the average per-user SE in regards to the number of LIS-units under the constraint of overall surface area in Fig. 6b.
From this figure, we see that OC can offer nearly stable gain of 2\,bits/Hz/s and 5\,bits/Hz/s with different $M$ for $R=5$\,m and $R=10$\,m, respectively, which indicates that OC benefits more for the scenarios with more users.
More importantly, it can be seen that the average per-user SE decreases monotonically with increasing the number of LIS-units. This is due to our association scheme such that each LIS-unit is assigned to one particular user. Therefore, under the association scheme used in this paper, the optimal number of LIS-units should be the number of users.

{We also show the impact of the order of PC implementation and OC implementation on the per-user SE in Fig. 7, where ``OC-PC'' represents the cases that the OC is done before the PC, and ``PC-OC'' represents the cases that the PC is done before the OC.
Firstly, it can be observed that the order of PC implementation and OC implementation has a significant impact on the achievable SE. The per-user spectral efficiency with ``OC-PC'' outperforms that with ``PC-OC'' for both cases of $K=5$ and $K=20$. For example, at $K=5$, compared to the ``PC-OC'' approach, the ``OC-PC'' approach can offer $2$~bits/Hz/s and $5$~bits/Hz/s SE gains regarding the median and 95\%-likely per-user SE, respectively. The performance gap increases when $K$ increases. The above result is reasonable since if the OC is done after the PC, then there will be a high probability that the change of the orientation may entirely modify the interference from the users. This compromises the benefits of PC.
}
%Firstly, it can be observed that the sequence of PC and OC has a significant impact of achievable SE. The per-user rate in D-LIS with ``OC-PC'' outperforms that with ``PC-OC'' for both 5 and 20 user cases.
%This phenomenon makes sense since the proposed suboptimal OC is aiming to reduce the interference caused by the nearest user to zero. Therefore, by performing OC ahead of PC, the results of PC algorithm is obtained with zero interference from nearest user as a premise. If changing the order of PC and OC, there is high probability that the change of the orientation may entirely modify the interference from each user making the PC becomes meaningless.
%Moreover, more SE gain can be obtained for more user cases. Compared with SE gain for $K=5$, the the median and 95\%-likely  per-user SE gain increases to 9 bits/Hz/s and 10 bits/Hz/s, respectively.
%The simulation indicates that `OC-PC'' can provide more stable and effective SE performance in a more complex interference environment.

We now look at the performance of proposed resource allocation algorithms in terms of minimum user SE. In Fig. 8, we compare the CDF of per-user SE evaluated by applying either PC or OC and both of the algorithms.
By noting that the proposed max-min PC and OC are available as well for C-LIS, we include the performance of C-LIS after PC and OC as benchmark.
Firstly, a key observation is that the proposed max-min PC and OC schemes increase remarkably the median and 95\%-likely per-user SE, and with these schemes, D-LIS significantly outperforms C-LIS, e.g., about 7\,bits/Hz/s and 10\,bits/Hz/s gain can be observed for the median and 95\%-likely per-user SE.
Moreover, both resource allocation algorithms offer more SE gain for the higher number of users cases. The most illustrative example is that the 95\%-likely per-user SE increases over 6 times with $K=20$, while such gain reduces remarkably when $K=20$.
In addition, we find that, with both resource allocation algorithms applied at LIS, the CDF performance of $K=20$ nearly approaches to that of $K=5$. This result implies that the proposed algorithms are greatly effective, and can make the distributed LIS deployment a very viable solution for future wireless networks.

{Finally, we compare the performance of the proposed D-LIS, C-LIS and conventional mMIMO in microwave band and millimeter wave (mmWave) band.
Fig. 9a and Fig. 9b show the CDFs of the achievable SE for C-LIS, D-LIS and conventional mMIMO at 2\,GHz and 50\,GHz, respectively. We set the number of antennas at the BS for conventional mMIMO as $1024$. For fairness of comparison, mMIMO deploys maximum-ratio combining at the BS and only LoS channels are considered in the system.
From Fig. 9a, it can be observed that D-LIS outperforms other architectures in microwave band for both $K=5$ and $K=20$ scenarios. The benefit comes from the macro-diversity gain obtained in D-LIS systems. Moreover, by comparing the performance of C-LIS and mMIMO, we find that when the number of users is small, C-LIS performs as well as  mMIMO. When the number of users increases, C-LIS showcases its superiority against conventional mMIMO.
We further illustrate the feasibility of LIS architecture in mmWave band in Fig. 9b. In contrast to the results in microwave band, C-LIS is superior to both D-LIS and conventional mMIMO for both $K=5$ and $K=20$ cases. The results are consistent with our theoretical analysis, where the LIS displays a strong ability of interference suppression at high frequency band. In addition, by comparing Fig. 9b and Fig. 9a, we barely observe any performance gain of D-LIS. This is due to the fact that the surface area of each D-LIS units is small which restricts the ability of interference suppression.}

\section{Conclusions}

In this paper, we have considered a \textbf{LIS-based} communication system, in which an LIS is viewed as an antenna array that can be used for transmission and reception. With MF used at the LIS, we have shown that the array gain and the spatial resolution of LIS architecture is proportional to its surface area and radius, respectively. Moreover, we have evaluated the relationship between the orientation of LIS and \emph{LIS response}, which indicates that the interference between users is highly dependent on the LIS orientation.
To give a full understanding of the \textbf{LIS-based} system, we have investigated the performance of C-LIS and D-LIS, and designed effective schemes to maximize the sum SE or maximize the minimum SE.
For C-LIS, by observing that the interference declines rapidly by increasing the surface area or frequency band, a searching based algorithm that maximizes the sum SE was proposed whose complexity is scaled down to the orientation domain.
Regarding D-LIS, we have designed a LSF-based user association scheme, an OC algorithm, and a max-min power control algorithm to fully showcase the potential of distributed systems in boosting diversity and coverage probability.
{The numerical results reveal that the proposed algorithms can very effectively enhance the system performance of both C-LIS and D-LIS. More importantly, we observe that D-LIS outperforms C-LIS in
microwave bands in terms of sum SE and minimum SE, while C-LIS shows its superiority in the mmWave band.  This result indicates that the operating frequency band is a critical factor that should be considered in the practical deployment of LIS.}

\appendices

\section{Proof for Proposition 1}

According to the analysis from (\ref{line_k}) to (\ref{delta_k}), the coefficient in (\ref{Coeffi_kk}) can be calculated as
\begin{align}
{\Sigma^{\mathcal{S}}_{kk'}} &= \iint_{\left( {x,y} \right) \in \mathcal{S}} {h_{k}^*\left( {x,y} \right){h_{k'}}\left( {x,y} \right)dxdy} \hfill \notag\\
&= {{\rm A}_{kk'}} \cdot \iint_{\left( {x,y} \right) \in \mathcal{S}} {{e^{j\kappa \left( {\Delta {d_k}\cos {\phi _k} - \Delta {d_{k'}}\cos {\phi _{k'}}} \right)}}dxdy} \hfill \notag\\
& = {{\rm A}_{kk'}} \cdot {\rm B} \left({R,\kappa,\chi _{kk'}}\right).
\end{align}
Since ${{\rm A}_{kk'}}$ is only dependent on the user's position, we thus focus on deriving ${\rm B} \left({R,\kappa,\chi _{kk'}}\right)$.
Substituting (\ref{delta_k}) into ${\rm B} \left({R,\kappa,\chi _{kk'}}\right)$, we get
\begin{align}\label{B_kk}
&{\rm B} \left({R,\kappa,\chi _{kk'}}\right) \notag \\
&= \iint_{\left( {x,y} \right) \in \mathcal{S}}{\cos \left( {\left( {\frac{{\kappa \cos {\phi _k}}}{{\sqrt {{{\tan }^2}{\alpha _k} + 1} }} - \frac{{\kappa \cos {\phi _{k'}}}}{{\sqrt {{{\tan }^2}{\alpha _{k'}} + 1} }}} \right)y} \right.}\notag\\
& \left. { - \left( {\frac{{\kappa \cos {\phi _k}\tan {\alpha _k}}}{{\sqrt {{{\tan }^2}{\alpha _k} + 1} }} - \frac{{\kappa \cos {\phi _{k'}}\tan {\alpha _{k'}}}}{{\sqrt {{{\tan }^2}{\alpha _{k'}} + 1} }}} \right)x} \right)dxdy.
\end{align}
Recalling the definition of $\tan{\alpha_k}$ and
$\cos{\phi_k}= \frac{\sqrt{x_k^2+y_k^2}}{d_k^c}$, it is easy to observe that
\begin{align}\label{angle2cor1}
\frac{{\cos {\phi _k}}}{{\sqrt {{{\tan }^2}{\alpha _k} + 1} }} = \frac{{{y_k}}}{{d_k^c}},
\end{align}
and
\begin{align}\label{angle2cor2}
\frac{{\cos {\phi _{k}}\tan {\alpha _{k}}}}{{\sqrt {{{\tan }^2}{\alpha _{k}} + 1} }} = \frac{{ {x_k}}}{{d_k^c}}.
\end{align}
Therefore, according to \emph{Definition} 1, we can simplify the above integration as
\begin{align}\label{B_kk_1}
{\rm B} \left(\!{R,\!\kappa,\!\chi _{kk'}}\!\right)\! = \! \iint_{\left( {x,y} \right) \in \mathcal{S}} {\cos \left( {\kappa \left( {{\eta _{kk'}}y \!-\! {\xi _{kk'}}x} \right)} \right)\!dxdy}.
\end{align}
%where
%\begin{align}\label{}
%{\eta _{{\rm Los},kk'}} = \frac{{\kappa \cos {\phi _k}}}{{\sqrt {{{\tan }^2}{\alpha _k} + 1} }} - \frac{{\kappa \cos {\phi _{k'}}}}{{\sqrt {{{\tan }^2}{\alpha _{k'}} + 1} }},
%\end{align}
%and
%\begin{align}\label{}
%{\xi _{{\rm Los},kk'}} = \frac{{\kappa \cos {\phi _k}\tan {\alpha _k}}}{{\sqrt {{{\tan }^2}{\alpha _k} + 1} }} - \frac{{\kappa \cos {\phi _{k'}}\tan {\alpha _{k'}}}}{{\sqrt {{{\tan }^2}{\alpha _{k'}} + 1} }}.
%\end{align}
Further, to obtain ${\rm B} \left({R,\kappa,\chi _{kk'}}\right)$ with a circular LIS, we transform the integration in (\ref{B_kk_1}) into polar coordinates, and the original integration is equivalent to
\begin{align}\label{B_kk_integ}
{\rm B} \left({R,\kappa,\chi _{kk'}}\right)&\mathop  = \limits^{(a)} 2\pi \int_0^R {r{J_0}\left( {r\kappa{{\chi _{kk'}}} } \right)dr}\notag\\
& \mathop  = \limits^{(b)}2\pi R\frac{{{J_1}\left( {R\kappa {\chi _{kk'}}} \right)}}{{\kappa {\chi _{kk'}}}},
\end{align}
%\begin{align}\label{B_kk_integ}
%{\rm B} \left({R,\kappa,\chi _{kk'}}\right)&= \int_0^R {\int_0^{2\pi } r{\cos \left( { r\kappa\left( {{\eta _{kk'}}\sin \vartheta  - {\xi _{kk'}}\cos \vartheta } \right)} \right)d\vartheta dr} }  \notag\\
%&\mathop  = \limits^{(a)} 2\pi \int_0^R {r{J_0}\left( {r\kappa{{\chi _{kk'}}} } \right)dr}\notag\\
%& \mathop  = \limits^{(b)}2\pi R\frac{{{J_1}\left( {R\kappa {\chi _{kk'}}} \right)}}{{\kappa {\chi _{kk'}}}},
%\end{align}
where $(a)$ is obtained by substituting (\ref{chi_kk}) into the integral, while $(b)$ can be obtained via \cite[Eq (6.521.1)]{gradshteyn2007ryzhik} with $J_{-1}(x)=- J_{1}(x)$ .

\section{Proof for Proposition 3}

Since the LIS unit can only adjust its angle along the $y$-axis, we draw the schematic diagram of LIS on the $xz$-plane, as shown in Fig. 3. Note that the variation of angle will not effect the distance between the user and LIS unit, but will create a $x^{\vartheta}z^{\vartheta}$-axis coordinate system which forces us to evaluate the user's coordinates in the new system. As shown in Fig. 4, we can obtain the user's coordinates as
\begin{align}\label{}
\left\{ {\begin{array}{*{20}{l}}
  {x_k^\vartheta  = \frac{{{x_k}}}{{\cos \vartheta }} + \sin \vartheta \left( {{z_k} - {x_k}\tan \vartheta } \right),} \\
  {y_k^\vartheta  = {y_k},} \\
  {z_k^\vartheta  = \cos \vartheta \left( {{z_k} - {x_k}\tan \vartheta } \right).}
\end{array}} \right.
\end{align}
Substituting the new coordinates into (\ref{eta_kk}) and (\ref{xi_kk}), we get
\begin{equation}\label{xi_theta_kk}
\xi _{kk'}^\vartheta  = {\xi _{kk'}},
\end{equation}
and
\begin{align}\label{eta_theta_kk}
\eta _{kk'}^\vartheta
&= \eta _{kk'}\left( {\frac{1}{{\cos \vartheta }} - \sin \vartheta \tan \vartheta } \right) + \zeta _{kk'}\sin \vartheta \hfill \notag\\
&= \eta _{kk'}\cos \vartheta + \zeta _{kk'}\sin \vartheta.
\end{align}
The square of $\eta _{kk'}^\vartheta $ then equals
\begin{align}\label{eta_theta_kk_2}
{\left(\! {\eta _{kk'}^\vartheta } \!\right)^2}\! = \!\eta _{kk'}^2{\cos ^2}\vartheta \! + \!{\zeta _{kk'}^2}{\sin ^2}\vartheta\!+ \!2{\eta _{kk'}}{\zeta _{kk'}}\cos \vartheta \sin \vartheta.
\end{align}
Then, substituting (\ref{xi_theta_kk}) and (\ref{eta_theta_kk_2}) into (\ref{chi_kk}), we complete the proof.

\section{Proof for Property 3}

The first-order derivative of ${\left( {\eta _{kk'}^{\vartheta}} \right)^2}$ with respect to ${\vartheta}$ is given as
\vspace{-0.1cm}
\begin{align}\label{deriv_eta_kk}
&{\partial {{\left( {\eta _{kk'}^{{\vartheta}}} \right)}^2}}/{\partial {\vartheta}}\notag\\
&\!=\! 2\!\left(\! {\zeta _{kk'}^2\! -\! \eta _{kk'}^2} \!\right)\sin {\vartheta}\cos {\vartheta}\! +\! 2{\eta _{kk'}}{\zeta _{kk'}}\!\left( {{{\cos }^2}{\vartheta} \!- \!{{\sin }^2}{\vartheta}} \!\right).
\vspace{-0.1cm}
\end{align}
By letting the result equal to zero, we have
\vspace{-0.1cm}
\begin{align}\label{equals_kk}
\frac{{\tan {\vartheta }}}{{1 - {{\tan }^2}{\vartheta}}} = \frac{{{\eta _{kk'}}{\zeta _{kk'}}}}{{\eta _{kk'}^2 - \zeta _{kk'}^2}}.
\end{align}
Recalling that
\vspace{-0.1cm}
\begin{equation}\label{}
\frac{{\tan {\vartheta}}}{{1 - {{\tan }^2}{\vartheta}}} = \frac{1}{2}\tan 2{\vartheta},
\end{equation}
and note that $\tan 2{\vartheta}$ is periodic in ${\vartheta}$ with period $\frac{\pi}{2}$, it is obvious that ${{\hat \vartheta }}$ has four solutions in the range $[{\pi},{\pi}]$.
Therefore, the four solutions can be expressed as %(\ref{var_theta_solu}).
\begin{equation}\label{}
\left\{ \begin{gathered}
  {{\hat \vartheta }_{1}} = \frac{1}{2}\arctan \frac{{2{\eta _{kk'}}{\zeta _{kk'}}}}{{\eta _{kk'}^2 - \zeta _{kk'}^2}}, \hfill \\
  {{\hat \vartheta }_{n}} = {{\hat \vartheta }_{1}} + \left(n-1\right)\frac{\pi }{2},\;\;\;n=2,3,4. \hfill \\
\end{gathered}  \right.
\end{equation}

To further determine the minimum and maximum value of ${\left( {\eta _{kk'}^{\vartheta}} \right)^2}$, we introduce a new parameter $v$, and treat $\vartheta$ as a function
\begin{equation}\label{func_varth}
{\vartheta}\left( v \right) = \frac{1}{2}\arctan v.
\end{equation}
By leveraging the following properties of trigonometric functions
\begin{align}
\label{sincos}&{\sin} {\vartheta}\left( v \right)\cos {\vartheta}\left( v \right)= \frac{v}{{2\sqrt {{v^2} + 1} }}, \hfill \\
\label{sin2}&{\sin ^2}{\vartheta}\left( v \right)= \frac{{1}}{{2}} - \frac{1}{{{2}\sqrt {{v^2} + 1} }},\hfill \\
\label{cos2}&{\cos ^2}{\vartheta}\left( v \right) = \frac{{1}}{{2}} + \frac{1}{{{2}\sqrt {{v^2} + 1} }},
\end{align}
we can transform (\ref{deriv_eta_kk}) as a function of $v$, and is given as
%\vspace{-0.1cm}
\begin{equation}\label{simply_v}
(\ref{deriv_eta_kk}) =\frac{{\left( {\zeta _{kk'}^2 - \eta _{kk'}^2} \right)v + 2{\eta _{kk'}}{\zeta _{kk'}}}}{{\sqrt {{v^2} + 1} }}.
\vspace{-0.2cm}
\end{equation}
It is clear that when $v < \frac{{2{\eta _{kk'}}{\zeta _{kk'}}}}{{\eta _{kk'}^2 - \zeta _{kk'}^2}}$, $(\ref{simply_v}) <0$, whilst when $v > \frac{{2{\eta _{kk'}}{\zeta _{kk'}}}}{{\eta _{kk'}^2 - \zeta _{kk'}^2}}$, $(\ref{simply_v}) >0$. Along with the fact that ${\vartheta}\left( v \right)$ increases monotonically with $v$, the solution ${{\hat \vartheta }_{1}}$ is the minimum point of $\left(\eta _{kk'}^{\vartheta}\right)^2$.
Then, considering the properties that $\sin \left( {{\vartheta } + \frac{\pi }{2}} \right) = \cos {\vartheta}$ and $\cos \left( {{\vartheta} + \frac{\pi }{2}} \right) =  - \sin {\vartheta }$, we obtain the solution ${{\hat \vartheta }_{3}}$ as another minimum point, while ${{\hat \vartheta }_{2}}$ and ${{\hat \vartheta }_{4}}$ are the maximum points.
Substituting (\ref{var_theta_solu}, \ref{sincos}-\ref{cos2}) into (\ref{eta2}) and after some manipulations, we can obtain ${\left( {\eta _{kk'}^{{\vartheta}}} \right)^2}$ equals to $0$ and $\eta _{kk'}^2 + \zeta _{kk'}^2$ for $\vartheta = {{\hat \vartheta }_{2}}$, and $\vartheta = {{\hat \vartheta }_{1}}$ respectively. Recalling (\ref{chi_vartheta}), we complete (\ref{range_chi}).

Clearly, the minimum value of ${{\rm{\tilde B}}( {{R},\kappa ,{\chi^\vartheta _{kk'}}} )}$ is either $0$ or the smaller value between ${{\rm{\tilde B}}( {{R},\kappa ,\chi _{kk'}^{\hat\vartheta_1 }})}$ and ${{\rm{\tilde B}}( {{R},\kappa ,\chi _{kk'}^{\hat\vartheta_2 }})}$. We therefore arrive at (\ref{min_norm_B}).
%Since ${{\rm{\tilde B}}( {{R},\kappa ,{\chi^\vartheta _{kk'}}} )}$ is an even function of the coefficient $\chi^\vartheta_{kk'}$, we hence only consider the positive cases, and simply obtain the minimum value ${{\rm{\tilde B}}( {{R},\kappa ,|{\xi _{kk'}}|} )}$.
If ${{\rm{\tilde B}}\left( {{R},\kappa ,{\chi _{kk'}^{\vartheta}}} \right)}$ has a zero point in the range of $[-\pi,\pi]$, based on the property of Bessel function, ${\chi ^{\vartheta}_{kk'}}$ should satisfy
\vspace{-0.1cm}
\begin{equation}\label{App_zero_j1}
{\chi ^{\vartheta}_{kk'}}=\frac{j_{1,n}}{R\kappa},\;\;n \in {\mathbb{N}_ + }.
\vspace{-0.1cm}
\end{equation}
Then, substituting (\ref{sincos}-\ref{cos2},\ref{App_zero_j1}) into (\ref{eta2}), the corresponding ${\vartheta}$ equals
\vspace{-0.1cm}
\begin{equation}\label{}
\vartheta = \frac{1}{2}\arctan {\bar v},
\vspace{-0.1cm}
\end{equation}
where ${\bar v}$ is the result of the following equation
\vspace{-0.1cm}
\begin{align}\label{Equ_v}
\left( {4\eta _{kk'}^2\zeta _{kk'}^2\! -\! c} \right){v^2}\! + \!{\text{4}}{\eta _{kk'}}{\zeta _{kk'}}\left( {\eta _{kk'}^2 \!- \!\zeta _{kk'}^2} \right)v &\!+ \!{\left( {\eta _{kk'}^2 \!- \!\zeta _{kk'}^2} \right)^2} \notag\\
&= c^2 .
\vspace{-0.1cm}
\end{align}
%with
%\begin{equation}\label{}
%c = 2{\left( {\frac{{{j_{1,n}}}}{{R\kappa }}} \right)^2} - \xi _{kk'}^{\text{2}} - \varpi _{kk'}^2.
%\end{equation}
Note that (\ref{Equ_v}) is a general quadratic equation, and the result can be solved as in (\ref{solve_v}).

\ifCLASSOPTIONcaptionsoff
  \newpage
\fi

\end{document}